%% file: ms.tex
\documentclass[longauth]{aa}
\usepackage{showyourwork}

\usepackage{graphicx}
\usepackage{txfonts,textcomp}
\usepackage{xcolor}
\usepackage{listings}
\usepackage{import}
\usepackage{fancyvrb}
\usepackage{color}
\usepackage[utf8]{inputenc}
\usepackage{hyperref}
\usepackage[autostyle]{csquotes}
\usepackage{siunitx}

\usepackage[switch]{lineno}

\input{code-examples/minted.tex}

\newcommand{\code}[1]{\texttt{#1}}

\newcommand{\ReadthedocsUrl}{\url{https://readthedocs.org/}\xspace}
\newcommand{\TravisUrl}{\url{https://www.travis-ci.org/}\xspace}

\newcommand{\astropy}{Astropy\xspace}
\newcommand{\gammapy}{Gammapy\xspace}
\newcommand{\scipy}{Scipy\xspace}
\newcommand{\numpy}{Numpy\xspace}
\newcommand{\iminuit}{iminuit\xspace}
\newcommand{\sherpa}{Sherpa\xspace}
\newcommand{\agnpy}{Agnpy\xspace}
\newcommand{\matplotlib}{Matplotlib\xspace}

\newcommand{\hess}{H.E.S.S.\xspace}
\newcommand{\hawc}{HAWC\xspace}
\newcommand{\veritas}{VERITAS\xspace}
\newcommand{\magic}{MAGIC\xspace}
\newcommand{\astri}{ASTRI\xspace}

\newcommand{\iacts}{IACTs\xspace}
\newcommand{\cta}{CTA\xspace}

\newcommand{\irf}{IRF\xspace}
\newcommand{\irfs}{IRFs\xspace}
\newcommand{\fermi}{Fermi-LAT\xspace}
\newcommand{\gammaray}{$\gamma$-ray\xspace}
\newcommand{\gammarays}{$\gamma$-rays\xspace}
\newcommand{\gadf}{GADF\xspace}
\newcommand{\milagro}{MILAGRO\xspace}
\newcommand{\github}{GitHub\xspace}

\begin{document}

\title{Gammapy: A Python package for gamma-ray astronomy}
\titlerunning{The Gammapy library, v1.0}
\authorrunning{The Gammapy project}

\author{
        Axel Donath \thanks{Mail inquiries to:
        \href{mailto:GAMMAPY-COORDINATION-L@IN2P3.FR}{GAMMAPY-COORDINATION-L@IN2P3.FR}} \inst{\ref{inst:001}} \and
        Régis Terrier \inst{\ref{inst:002}} \and
        Quentin Remy \inst{\ref{inst:003}} \and
        Atreyee Sinha \inst{\ref{inst:004}} \and
        Cosimo Nigro \inst{\ref{inst:005}} \and
        Fabio Pintore \inst{\ref{inst:006}} \and
        Bruno Khélifi \inst{\ref{inst:002}} \and
        Laura Olivera-Nieto \inst{\ref{inst:003}}\and
        Jose Enrique {Ruiz} \inst{\ref{inst:007}} \and
        Kai Brügge \inst{\ref{inst:008},\ref{inst:009}} \and
        Maximilian Linhoff \inst{\ref{inst:009}} \and
        Jose Luis {Contreras} \inst{\ref{inst:004}} \and
        Fabio Acero \inst{\ref{inst:010}} \and
        Arnau Aguasca-Cabot \inst{\ref{inst:011}, \ref{inst:012}, \ref{inst:013}, \ref{inst:014}} \and
        David Berge \inst{\ref{inst:015}, \ref{inst:016}} \and
        Pooja Bhattacharjee \inst{\ref{inst:017}} \and
        Johannes Buchner \inst{\ref{inst:018}} \and
        Catherine Boisson \inst{\ref{inst:019}} \and
        David {Carreto Fidalgo} \inst{\ref{inst:020}} \and
        Andrew Chen \inst{\ref{inst:021}} \and
        Mathieu {de Bony de Lavergne} \inst{\ref{inst:017}} \and
        José Vinícius {de Miranda Cardoso}\inst{\ref{inst:022}} \and
        Christoph Deil \inst{\ref{inst:003}} \and
        Matthias Fü{\ss}ling \inst{\ref{inst:023}} \and
        Stefan Funk \inst{\ref{inst:024}} \and
        Luca Giunti \inst{\ref{inst:002}} \and
        Jim Hinton \inst{\ref{inst:003}} \and
        Léa Jouvin \inst{\ref{inst:025}} \and
        Johannes King \inst{\ref{inst:026}, \ref{inst:003}} \and
        Julien Lefaucheur \inst{\ref{inst:027}, \ref{inst:002}} \and
        Marianne Lemoine-Goumard \inst{\ref{inst:028}} \and
        Jean-Philippe Lenain \inst{\ref{inst:029}} \and
        Rub{\'e}n L{\'o}pez-Coto \inst{\ref{inst:007}} \and
        Lars Mohrmann \inst{\ref{inst:003}} \and
        Daniel Morcuende \inst{\ref{inst:004}} \and
        Sebastian Panny \inst{\ref{inst:032}} \and
        Maxime Regeard \inst{\ref{inst:002}} \and
        Lab Saha \inst{\ref{inst:004}} \and
        Hubert Siejkowski \inst{\ref{inst:030}} \and
        Aneta Siemiginowska \inst{\ref{inst:001}} \and
        Brigitta M. {Sipőcz} \inst{\ref{inst:031}} \and
        Tim Unbehaun \inst{\ref{inst:024}} \and
        Christopher {van Eldik} \inst{\ref{inst:024}} \and
        Thomas Vuillaume \inst{\ref{inst:017}} \and
        Roberta Zanin \inst{\ref{inst:023}}
}

\institute{
        Center for Astrophysics | Harvard and Smithsonian \label{inst:001} \and
        Université de Paris Cité, CNRS, Astroparticule et Cosmologie, F-75013 Paris, France  \label{inst:002} \and
        Max-Planck-Institut für Kernphysik, P.O. Box 103980, D 69029 Heidelberg, Germany \label{inst:003} \and
        IPARCOS Institute and EMFTEL Department, Universidad Complutense de Madrid, E-28040 Madrid, Spain \label{inst:004} \and
        Institut de Física d'Altes Energies (IFAE), The Barcelona Institute of Science and Technology, Campus UAB, Bellaterra, 08193 Barcelona, Spain \label{inst:005} \and
        INAF/IASF Palermo, Via U. La Malfa, 153, 90146 Palermo PA \label{inst:006} \and
        Instituto de Astrofísica de Andalucía-CSIC, Glorieta de la Astronomía s/n, 18008, Granada, Spain \label{inst:007} \and
        Point8 GmbH \label{inst:008} \and
        Astroparticle Physics, Department of Physics, TU Dortmund University, Otto-Hahn-Str. 4a, D-44227 Dortmund \label{inst:009} \and
        Université Paris-Saclay, Université Paris Cité, CEA, CNRS, AIM, F-91191 Gif-sur-Yvette, France \label{inst:010} \and
        Departament de Física Quàntica i Astrofísica (FQA), Universitat de Barcelona (UB),  c. Martí i Franqués, 1, 08028 Barcelona, Spain \label{inst:011} \and
        Institut de Ciències del Cosmos (ICCUB), Universitat de Barcelona (UB), c. Martí i Franqués, 1, 08028 Barcelona, Spain \label{inst:012} \and
        Institut d'Estudis Espacials de Catalunya (IEEC), c. Gran Capità, 2-4, 08034 Barcelona, Spain \label{inst:013} \and
        Departament de Física Quàntica i Astrofísica, Institut de Ciències del Cosmos, Universitat de Barcelona, IEEC-UB, Martí i Franquès, 1, 08028, Barcelona, Spain \label{inst:014} \and
        Deutsches Elektronen-Synchrotron (DESY), D-15738 Zeuthen, Germany \label{inst:015} \and
        Institute of physics, Humboldt-University of Berlin, D-12489 Berlin, Germany \label{inst:016} \and
        Université Savoie Mont Blanc, CNRS, Laboratoire d’Annecy de Physique des Particules - IN2P3, 74000 Annecy, France \label{inst:017} \and
        Max Planck Institute for extraterrestrial Physics, Giessenbachstrasse, 85748 Garching, Germany \label{inst:018} \and
        Laboratoire Univers et Théories, Observatoire de Paris, Université PSL, Université Paris Cité, CNRS, F-92190 Meudon, France \label{inst:019} \and
        Max Planck Computing and Data Facility, Gießenbachstraße 2, 85748 Garching \label{inst:020} \and
        School of Physics, University of the Witwatersrand, 1 Jan Smuts Avenue, Braamfontein, Johannesburg, 2050 South Africa \label{inst:021} \and
        The Hong Kong University of Science and Technology, Department of Electronic and Computer Engineering \label{inst:022} \and
        Cherenkov Telescope Array Observatory gGmbH (CTAO gGmbH) Saupfercheckweg 1 69117 Heidelberg \label{inst:023} \and
        Erlangen Centre for Astroparticle Physics (ECAP), Friedrich-Alexander-Universität Erlangen-Nürnberg, Nikolaus-Fiebiger Strasse 2, 91058 Erlangen, Germany \label{inst:024} \and
        IRFU, CEA, Université Paris-Saclay, F-91191 Gif-sur-Yvette, France \label{inst:025}
        Bruno-Bauer-Straße 22, 12051 Berlin  \label{inst:026} \and
        Meteo France International \label{inst:027} \and
        Université Bordeaux, CNRS, LP2I Bordeaux, UMR 5797 \label{inst:028} \and
        Sorbonne Université, Université Paris Diderot, Sorbonne Paris Cité, CNRS/IN2P3, Laboratoire de Physique Nucléaire et de Hautes Energies, LPNHE, 4 Place Jussieu, F-75252 Paris, France \label{inst:029} \and
        Academic Computer Centre Cyfronet, AGH University of Science and Technology, Krakow, Poland \label{inst:030} \and
        Caltech/IPAC, MC 100-22, 1200 E. California Boulevard, Pasadena, CA 91125 USA \label{inst:031} \and
        Institut für Astro- und Teilchenphysik, Leopold-Franzens-Universität Innsbruck, A-6020 Innsbruck, Austria \label{inst:032}
}

\abstract
        {
                Traditionally, TeV-\gammaray astronomy has been conducted
                by experiments employing proprietary data and analysis software.
                However, the next generation of \gammaray instruments,
                such as the Cherenkov Telescope Array Observatory (CTAO), will be operated as open observatories.
                Alongside the data, they will also make the associated software tools available to a wider community.
    This necessity prompted the development of open, high-level, astronomical software customized for high-energy astrophysics.
        }
        {
                In this article, we present \gammapy, an open-source Python package for the analysis of astronomical \gammaray data,
                and illustrate the functionalities of its first long-term-support release, version 1.0.
                Built on the modern Python scientific ecosystem, \gammapy provides a uniform platform for reducing and
                modeling data from different \gammaray instruments for many analysis scenarios.
    \gammapy complies with several well-established data conventions in high-energy astrophysics, providing serialized data products that
                are interoperable with other software packages.
        }
        {
                Starting from event lists and instrument response functions,
                \gammapy provides functionalities to reduce these data by binning them in energy and sky coordinates.
                Several techniques for background estimation are implemented in the package to handle the residual hadronic background affecting \gammaray instruments.
                After the data are binned, the flux and morphology of one or more \gammaray sources can be estimated
                using Poisson maximum likelihood fitting
                and assuming a variety of spectral, temporal, and spatial models.
                Estimation of flux points, likelihood profiles, and light curves is also supported.
        }
        {
                After describing the structure of the package, we show, using publicly available gamma-ray data,
                the capabilities of \gammapy in multiple traditional and novel \gammaray analysis scenarios,
                such as spectral and spectro-morphological modeling and estimations of a spectral energy
                distribution and a light curve.
                Its flexibility and its power are displayed in a final multi-instrument example,
                where datasets from different instruments, at different stages of data reduction,
                are simultaneously fitted with an astrophysical flux model.
        }{}


\maketitle

\section{Introduction}
\label{sec:introduction}
\begin{figure*}[t]
        \centering
        \includegraphics[width=\textwidth]{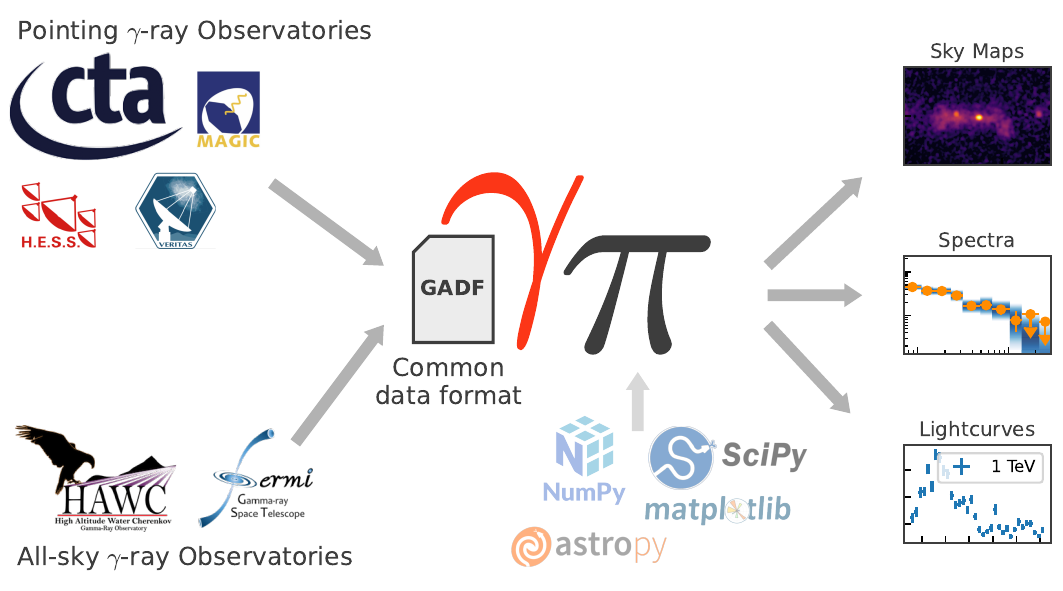}
        \caption{
                Core idea and relation of \gammapy to different \gammaray instruments
                and the gamma astro data format (GADF). The top left shows the
                group of current and future pointing instruments based on the 
                imaging atmospheric Cherenkov technique (IACT). This includes
                instruments such as the Cherenkov telescope array observatory (CTAO),
                the high energy stereoscopic system (H.E.S.S.), the
                major atmospheric gamma imaging Cherenkov telescopes (MAGIC),
                and the very energetic radiation imaging telescope array system (VERITAS).
                The lower left shows the group of all-sky instruments such as the
                Fermi large area telescope (Fermi-LAT) and the high altitude
                water Cherenkov observatory (HAWC). The calibrated data of all those
                instruments can be converted and stored into the common GADF data format, which \gammapy can read.
                The \gammapy package is a community-developed project that
                provides a common interface to the data and analysis of all
                these \gammaray instruments, allowing users to easily combine data from
                different instruments and perform joint analyses.
                \gammapy is built on the scientific Python ecosystem, and the required dependencies
                are shown below the \gammapy logo.
        }
        \label{fig:big_picture}

\end{figure*}
%
Modern astronomy offers the possibility to observe and study astrophysical sources across
all wavelengths. The \gammaray range of the electromagnetic spectrum 
provides us with insights into the most energetic processes in the Universe such as
those accelerating particles in the surroundings of black holes or remnants of
supernova explosions.

In general, \gammaray astronomy relies on the detection of individual 
photon events and reconstruction of their incident direction as well
as energy. As in other branches of astronomy, this can be
achieved by satellite as well as ground-based \gammaray instruments.
Space-borne instruments such as the Fermi Large Area Telescope (LAT)
rely on the pair-conversion effect to detect \gammarays and track
the positron-electron pairs created in the detector to reconstruct the incident direction 
of the incoming \gammaray. The energy of the photon is estimated using a calorimeter
at the bottom of the instrument. The energy range of instruments such as \fermi \citep{Atwood2009}, referred to as
\enquote{high energy} (HE), goes approximately from tens of Megaelectronvolts to hundreds of Gigaelectronvolts. 

Ground-based instruments, instead, use Earth's atmosphere as a particle detector, relying on the effect that
cosmic \gammarays interacting in the atmosphere create large cascades of secondary particles, so called \enquote{air showers}, that can be observed from the ground.
Ground-based \gammaray astronomy relies on the observation of these extensive air showers to estimate the
primary \gammaray photons' incident direction and energy.
These instruments operate in the so-called \enquote{very high energy} (VHE) regime,
covering the energy range from a few tens of Gigaelectronvolts up to Petaelectronvolts.
There are two main categories of ground-based instruments.

First there are imaging atmospheric Cherenkov telescopes (IACTs), which obtain images of the atmospheric showers
by detecting the Cherenkov radiation emitted by charged particles in the cascade and
they use these images to reconstruct the properties of the incident particle.
Those instruments have a limited field of view (FoV) and duty cycle, but
good energy and angular resolution.
        
Second there are water Cherenkov detectors (WCDs), that detect particles directly from the tail of the
shower when it reaches the ground. These instruments have a very
large FoV, and large duty-cycle, but a higher energy threshold and
lower signal-to-noise ratios compared to IACTs~\citep{2015CRPhy..16..610D}.

While \fermi data and analysis tools have been public since the early years
of the project \citep{Atwood2009}, ground-based \gammaray astronomy has been historically conducted
through experiments operated by independent collaborations, each relying
on their own proprietary data and analysis software developed as part of the
instrument. While this model has been successful so far, it does not
permit data from several instruments to be combined easily. This lack of
interoperability currently limits the full exploitation of the
available \gammaray data, especially in light of the fact that different instruments often have
complementary sky coverages, and the various detection
techniques have complementary properties in terms of the energy range covered,
duty cycle, and spatial resolution.

The Cherenkov Telescope Array Observatory (CTAO) will be the first ground-based
\gammaray instrument to be operated as an open observatory.
Its high-level data\footnote{The lowest reduction level of data published by \cta will be reconstructed event lists and corresponding instrument response functions.}
will be shared publicly after some proprietary period, along with the software required to analyze them.
The adoption of an open and standardized \gammaray data format, besides being a necessity for future observatories such as \cta,
will be extremely beneficial to the current generation of instruments, eventually allowing their data legacies to be exploited even
beyond the end of their scientific operations.

The usage of a common data format is facilitated by the remarkable similarity of the data reduction workflow of all \gammaray telescopes.
After data calibration, shower events are reconstructed and
a gamma-hadron separation is applied to reject cosmic-ray-initiated showers and build lists of \gammaray-like events.
The latter can then be used, taking into account the observation-specific instrument response functions (IRFs),
to derive scientific results, such as spectra, sky maps, or light curves.
Once the data is reduced to a list of events with reconstructed physical properties of the primary particle,
the information is independent of the data-reduction process, and, eventually, of the detection technique. This implies,
for example, that high-level data from IACTs and WCDs can be represented
with the same data model.
The efforts to create a common format usable by various instruments
converged in the so-called {data formats for \gammaray astronomy}
initiative~\citep{gadf_proc,gadf_universe}, abbreviated to
\texttt{gamma-astro-data-formats} (\gadf). This community-driven initiative proposes prototypical
specifications to produce files based on the flexible image transport system
(FITS) format~\citep{fits} encapsulating this high-level information. This is
realized by storing a list of \gammaray-like events with their reconstructed and observed
quantities such as energy, incident direction and arrival time, and a parameterization of
the IRFs associated with the event list data. 

In the past decade, {Python} has become extremely popular as a scientific programming language,
in particular in the field of data sciences. This success is
mostly attributed to the simple and easy to learn syntax, the ability to act as
a \enquote{glue} language between different programming languages, and last but not least
the rich ecosystem of packages and its open and supportive community \citep{Momcheva2015}.
In the subfield of computational astronomy, the \astropy project \citep{astropy} was created in 2012
to build a community-developed core Python package for astronomy.
It offers basic functionalities that astronomers of many fields need, such as representing
and transforming astronomical coordinates, manipulating physical quantities including units,
as well as reading and writing FITS files.

The \gammapy project was started following the model of \astropy, with the objective of building a common
software library for \gammaray data analysis \citep{gammapy_2015}. 
The core of the idea is illustrated in Figure~\ref{fig:big_picture}. The various \gammaray instruments
can export their data to a common format (\gadf) and then combine and analyze
them using a common software library.
The \gammapy package is an independent community-developed software project.
It has been selected to be the core library for the science analysis tools of \cta,
but it also involves contributors associated with other instruments.
The \gammapy package is built on the scientific Python ecosystem: it uses \numpy~\citep{numpy} for ND data
structures, \scipy~\citep{2020SciPy-NMeth} for numerical algorithms, \astropy~\citep{astropy} for
astronomy-specific functionality, \iminuit~\citep{iminuit} for numerical minimization,
and \matplotlib~\citep{matplotlib} for visualization.

With the public availability of the \gadf format specifications and the
\gammapy package, some experiments have started to make limited subsets of
their \gammaray data publicly available for testing and validating
\gammapy. For example, the \hess collaboration released a limited test
dataset (about 50 hours of observations taken between 2004 and 2008)
based on the \gadf format \citep{HESS_DR1} for data level 3 (DL3) \gammaray data.
This data release served as a basis for validation of open analysis tools, including \gammapy 
\cite[see e.g.,][]{Mohrmann2019}. Two observations of the Crab nebula have
been released by the \magic collaboration~\citep{magic_performance}.
Using these public data from \fermi, \hess, \magic, and additional observations
provided by FACT and \veritas, the authors of \cite{joint_crab} presented
a combined analysis of \gammaray data from different instruments for the first time.
Later the \hawc collaboration also released a limited test dataset of the Crab Nebula,
which was used to validate the \gammapy package in \cite{Olivera2022}.

The increased availability of public data that followed the definition of
a common data format, and the development of \gammapy as a community-driven open software,
led the way toward a more open science in the VHE \gammaray astronomy domain.
The adoption of \gammapy as science tools strengthens the commitment of the future \cta Observatory to the  findable, accessible, interoperable, and reusable (FAIR) principles \citep{FAIR16, FAIR22} 
that define the key requirements for open science.

In this article, we describe the general structure of the \gammapy package,
its main concepts, and organizational structure. We start in
Section~\ref{sec:gammaray-data-analysis} with a general overview
of the data analysis workflow in VHE \gammaray astronomy. Then we
show how this workflow is reflected in the structure of the \gammapy package 
in Section~\ref{sec:gammapy-package}, while also
describing the various sub-packages it contains. Section~\ref{sec:applications}
presents a number of applications, while Section~\ref{sec:gammapy-project}
finally discusses the project organization.

\section{Gamma-ray data analysis}
\begin{figure*}[t]
        \centering
        \includegraphics[width=1.\textwidth]{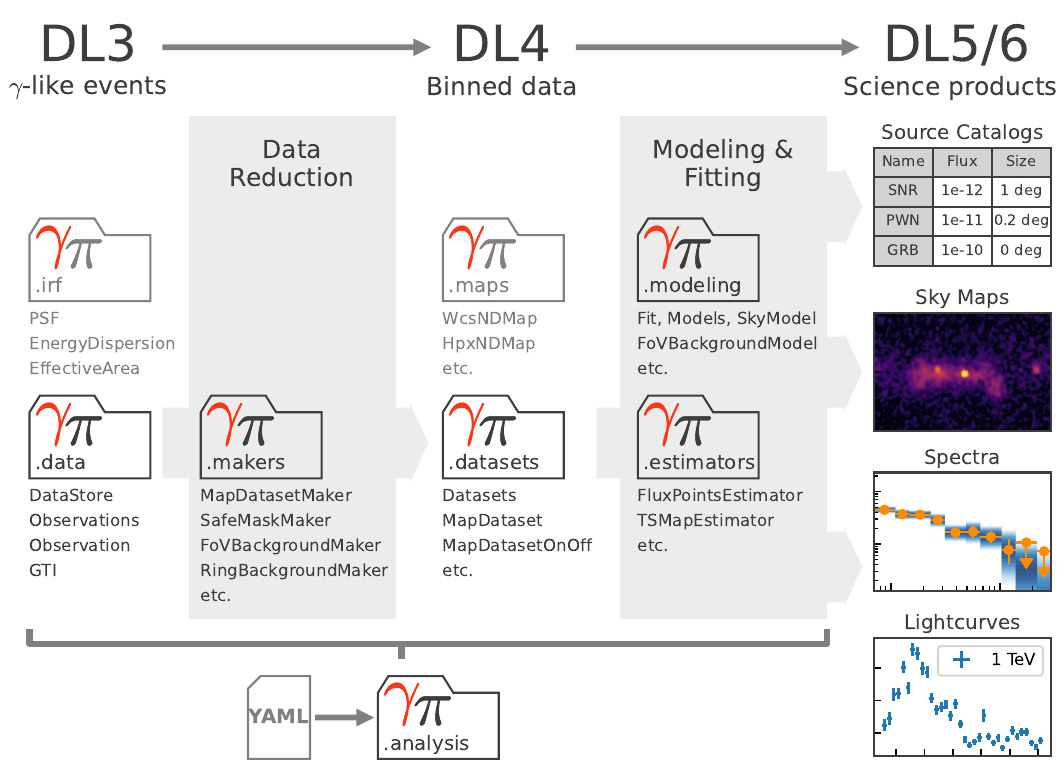}
        \caption{
                \gammapy sub-package structure and data analysis workflow. The top row
        defines the different levels of data reduction, from lists of \gammaray-like
        events on the left (DL3), to high-level scientific products
        (DL5) on the right. The direction of the data flow is illustrated with the
        gray arrows. The gray folder icons represent the different sub-packages
        in \gammapy and names given as the corresponding Python code suffix, e.g., 
                \code{gammapy.data}. Below each icon there is a list of the most
        important objects defined in the sub-package. The light gray folder
                icons show the sub-packages for the most fundamental data structures such 
                as maps and IRFs. The bottom of the figure shows the high-level analysis
                sub-module with its dependency on the YAML file format. 
    }
        \label{fig:data_flow}
\end{figure*}
\label{sec:gammaray-data-analysis}
The data analysis process in \gammaray astronomy is usually split into two stages.
The first one deals with the data processing from detector measurement, calibration, event
reconstruction and selection to yield a list of reconstructed \gammaray event candidates.
This part of the data reduction sequence, sometimes referred to as low-level analysis,
is usually very specific to a given observation technique and even to a given instrument.

The second stage, referred to as high-level analysis, deals with the extraction of physical
quantities related to \gammaray sources and the production of high-level science products
such as spectra, light curves and catalogs. The core of the analysis consists in predicting
the result of an observation by modeling the flux distribution of an astrophysical
object and pass it through a model of the instrument. The methods applied here are more
generic and are broadly shared across the field. The similarity in the high-level
analysis conceptionally also allow for easily combining data from multiple instruments.
This part of the data analysis is supported by \gammapy. 

\subsection{DL3: events and instrument response functions}
An overview of the typical steps in the high level analysis is shown in the upper
row of Figure~\ref{fig:data_flow}. The high level analysis starts at the DL3 
data level, where \gammaray data is represented as lists of \gammaray-like events and their
corresponding IRFs $R$ and ends at the DL5/6 data level, where the physically relevant 
quantities such as fluxes, spectra and light curves of sources have been derived. DL3 data are typically 
bundled into individual observations, corresponding to
stable periods of data acquisition. For IACT instruments, for which the \gadf
data model and \gammapy were initially conceived, this duration is typically 
$t_{\rm obs}=15 - 30\,{\rm min}$. Each observation is assigned a unique integer ID for reference.
The event list is just a simple table with one event per row and
the measured event properties as columns. These properties for example, include reconstructed
incident direction and energy, arrival time and reconstruction quality.

A common assumption for the instrument response is that it can be simplified as the product 
of three independent functions:

\begin{equation}
        \begin{split}
   R(p, E|p_{\rm true}, E_{\rm true}) = &~A_{\rm eff}(p_{\rm true}, E_{\rm true}) \\
        & \cdot PSF(p|p_{\rm true}, E_{\rm true})\\
    & \cdot E_{\rm disp}(E|p_{\rm true}, E_{\rm true})
        \end{split}
,\end{equation}

\noindent where
\begin{itemize}
\setlength\itemsep{1em}
\item $A_{\rm eff}(p_{\rm true}, E_{\rm true})$ is the effective collection area of the detector. It is the product
  of the detector collection area times its detection efficiency at true energy $E_{\rm true}$ and position $p_{\rm true}$.
\item $PSF(p|p_{\rm true}, E_{\rm true})$ is the point spread function (PSF). It gives the probability density of
  measuring a direction $p$ when the true direction is $p_{\rm true}$ and the true energy is $E_{\rm true}$.
  \gammaray instruments typically consider radial symmetry of the PSF. With this assumption the probability density 
  $PSF(\Delta p|p_{\rm true}, E_{\rm true})$ only depends on the angular separation between true
 and reconstructed direction defined by $\Delta p = p_{\rm true} - p$.  
\item $E_{\rm disp}(E|p_{\rm true}, E_{\rm true})$ is the energy dispersion. It gives the probability to
  reconstruct the photon at energy $E$ when the true energy is $E_{\rm true}$ and the true position $p_{\rm true}$.
  \gammaray instruments consider $E_{\rm disp}(\mu|p_{\rm true}, E_{\rm true})$, the probability density of the event
  migration, $\mu=\frac{E}{E_{\rm true}}$.
\end{itemize}

In addition to the instrument characteristics described above, there is also the instrumental
background, that results from hadronic events being misclassified as \gammaray events. These
events constitute a uniform background to the \gammaray events. For \fermi the residual hadronic background
is very small ($<1\%$) because of its veto layer and can often be neglected. For IACTs and WCDs in contrast,
the background can account for a very large part ($>95\%$) of the events and must be treated
accordingly in the analysis. As the background is very specific to the instrument, \gammapy typically 
relies on the background models provided with the DL3 data. The background usually only depends
on the reconstructed event position and energy $Bkg(p, E)$. However in general all IRFs depend on the
geometrical parameters of the detector, such as location of an event in the FoV or the elevation
angle of the incoming direction of the event. Consequently IRFs might be parameterized as functions
of detector specific coordinates as well.

The first step in \gammaray data analysis is the selection and extraction of a subset of
observations based on their metadata including information such as pointing direction,
observation time and observation conditions. All functionality related to representation, 
access and selection of DL3 data is available in the \code{gammapy.data} and \code{gammapy.irf}
sub-packages.

\subsection{From DL3 to DL4: data reduction}
The next step of the analysis is the data reduction, where all observation events and instrument
responses are filled into or projected onto a common physical coordinate system, defined by
a map geometry. The definition of the map geometry typically consists of a spectral dimension
defined by a binned energy axis and of spatial dimensions, which either define 
a spherical projection from celestial coordinates to a pixelized image space
or a single region on the sky.

After all data have been projected onto the same geometry, it is typically
required to improve the residual hadronic background estimate. As residual hadronic
background models can be subject to significant systematic uncertainties,
these models can be improved by taking into account actual data
from regions without known \gammaray sources. This includes methods 
such as the ring or the FoV background normalization techniques or background measurements
performed for example within reflected regions~\citep{Berge07}.

Data measured at the FoV or energy boundaries of the instrument are typically
associated with a systematic uncertainty in the IRF. For this reason this part 
of the data is often excluded from subsequent analysis by defining regions of
\enquote{safe} data in the spatial as well as energy dimension. All the code
related to reduce the DL3 data to binned data structures is contained in the
\code{gammapy.makers} sub-package.

\subsection{DL4: binned data structures}
The counts data and the reduced IRFs in the form of maps are bundled into datasets
that represent the fourth data level (DL4). These reduced datasets can be written to disk,
in a format specific to \gammapy to allow users to read them back at any time later
for modeling and fitting. Different variations of such datasets support different 
analysis methods and fit statistics. The datasets can be used to perform a joint-likelihood
fit, allowing one to combine different measurements from different observations,
but also from different instruments or event classes. They can also be used for binned
simulation as well as event sampling to simulate DL3 events data.

Binned maps and datasets, which represent a collection of binned
maps, are defined in the \code{gammapy.maps} and \code{gammapy.datasets}
sub-packages, respectively.

\subsection{From DL4 to DL5/6: modeling and fitting}
The next step is then typically to model the datasets using binned Poisson maximum likelihood fitting.
Assuming Poisson statistics per bin the log-likelihood of an observation is given by the Cash statistics\citep{Cash}:
\begin{equation}
        \mathcal{C} = 2 \sum_{i=0}^K N_{\mathrm{Pred, i}} - N_{\mathrm{Obs, i}} \log{N_{\mathrm{Pred, i}}} 
\end{equation}

Where the expected number of events for the observation is given by forward folding a source model through
the instrument response:
 \begin{equation}
         \begin{split}
                N_{\mathrm{Pred}}(p, E; \hat{\theta})\ {\rm d}p\ {\rm d}E =  &E_{\rm disp} \circledast \left[ PSF \circledast \left( A_{\rm eff} \cdot t_{\rm obs} \cdot \Phi(\hat{\theta}) \right) \right]\\
                                                & + Bkg(p, E) \cdot t_{\rm obs}
         \end{split}
 .\end{equation}

The equation includes the IRF components described in the previous section, as well as an analytical model to describe the
intensity of the radiation from \gammaray sources as a function of the energy, $E_{\rm true}$,
and of the position in the FoV, $p_{\rm true}$:
\begin{equation}
    \Phi(p_{\rm true}, E_{\rm true}; \hat{\theta}), [\Phi] = \si{TeV^{-1}.cm^{-2}.s^{-1}}
    \label{eq:general_source_model}
,\end{equation}

where $\hat{\theta}$ is a set of model parameters that can be adjusted in a fit.

Observations can be either modeled individually, or in a joint likelihood analysis.
In the latter case the total (joint) log-likelihood is given by the sum of the log-likelihoods
per observation $\mathcal{L} = \sum{i=0}^M \mathcal{C_i}$ for M observations. Other fit
statistics such as \emph{WStat}~\citep{WStat}, where the background is estimated from an independent
measurement or a simple $\chi^2$ for flux points are also often used
and are supported by \gammapy as well. Models can be simple analytical models or more complex
ones from radiation mechanisms of accelerated particle populations such as inverse Compton or $\pi^{0}$ decay.

Independently or subsequently to the global modeling, the data can be regrouped to compute
flux points, light curves and flux maps as well as significance maps in different energy bands.

Parametric models and all the functionality related to fitting is implemented in 
\code{gammapy.modeling} and \code{gammapy.estimators}, where the latter is used to
compute higher level science products such as flux and significance maps,
light curves or flux points.

 



\section{\gammapy package}
\label{sec:gammapy-package}
\subsection{Overview}
\label{ssec:overview}
The \gammapy package is structured into multiple sub-packages. 
Figure~\ref{fig:data_flow} also shows the overview of the different sub-packages and
their relation to each other. The definition
of the content of the different sub-packages follows mostly the stages of the
data reduction workflow described in the previous section. Sub-packages either
contain structures representing data at different reduction levels or algorithms
to transition between these different levels. In the following sections, we will
introduce all sub-packages and their functionalities in more detail. In the online 
documentation we also provide an overview of all the \gammapy sub-packages\footnote{\url{https://docs.gammapy.org/1.0/user-guide/package.html}}.

\subsection{gammapy.data}
\label{ssec:gammapy-data}
The \code{gammapy.data} sub-package implements the functionality to select,
read, and represent DL3 \gammaray data in memory. It provides the main user
interface to access the lowest data level. \gammapy currently only
supports data that is compliant with \code{v0.2} and \code{v0.3} of the \gadf data format.

A typical usage example is shown in Figure~\ref{fig*:minted:gp_data}.
First a \code{DataStore} object is created from the path of the data
directory. The directory contains an observation as well as a FITS HDU \footnote{Header Data Unit} 
index file which assigns the correct data and IRF FITS files and HDUs
to the given observation ID. The \code{DataStore}
object gathers a collection of observations and provides ancillary
files containing information about the telescope observation mode and the
content of the data unit of each file. The \code{DataStore} allows for
selecting a list of observations based on specific filters.

The DL3 level data represented by the \code{Observation} class consist
of two types of elements: first, the list of \gammaray events 
which is represented by the \code{EventList} class. Second, a set of
associated IRFs, providing the response of the system, typically
factorized in independent components as described in
Section~\ref{ssec:gammapy-irf}. The separate handling of event lists and IRFs
additionally allows for data from non-IACT \gammaray instruments to be read. For
example, to read \fermi data, the user can read separately their event list
(already compliant with the \gadf specifications) and then find the appropriate
IRF classes representing the response functions provided by \fermi, see
example in Section~\ref{ssec:multi-instrument-analysis}.

\begin{figure}
        \small
        \import{code-examples/generated/}{gp_data}
        \caption{
        Using \code{gammapy.data} to access DL3 level data with a \code{DataStore} object.
        Individual observations can be accessed by their unique integer observation id number.
        The actual events and instrument response functions can be accessed
        as attributes on the \code{Observation} object, such as \code{.events}
        or \code{.aeff} for the effective area information. The output
                of the code example is shown in Figure~\ref{fig:code_example_gp_data}.
    }
        \label{fig*:minted:gp_data}
\end{figure}

\subsection{gammapy.irf}
\label{ssec:gammapy-irf}
\begin{figure*}[t]
        \centering
        \includegraphics[width=1.\textwidth]{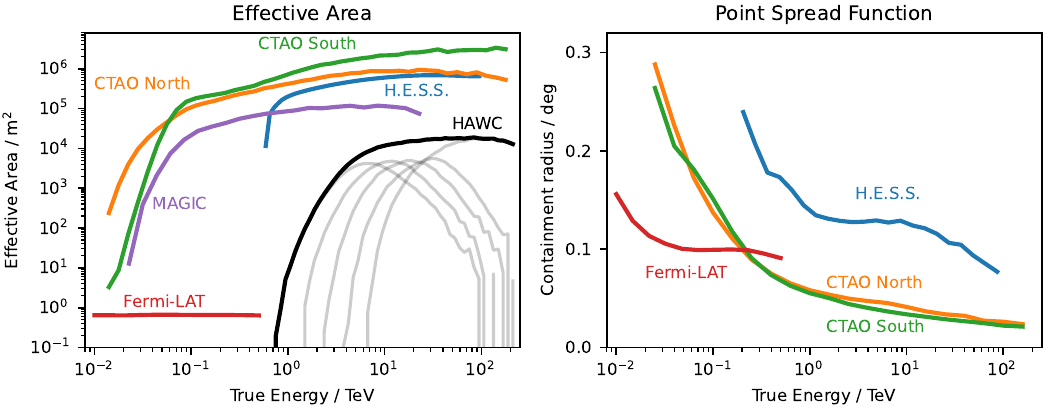}
        \caption{
                Using \code{gammapy.irf} to read and plot instrument response functions.
                The left panel shows the effective area as a function of energy for
                the \cta, \hess, \magic, \hawc and \fermi instruments. The right panel shows
                the $68\%$ containment radius of the PSF as a function of energy for the \cta, \hess
                and \fermi instruments. The \cta IRFs are from the \enquote{prod5} production for the {\it alpha} configuration of the south and north array. The \hess IRFs are from the DL3 DR1,
        using observation ID 033787. The \magic effective area is computed for a
        $20\,{\rm min}$ observation at the Crab Nebula coordinates. The
                \fermi IRFs use \enquote{pass8} data and are also taken at the position of the Crab Nebula.
                The \hawc effective area is shown for the event classes $N_{Hit}=5 - 9$ as light gray
                lines along with the sum of all event classes as a black line. The \hawc IRFs are taken from
                the first public release of events data by the \hawc collaboration. All IRFs do not correspond
                to the latest performance of the instruments, but still are representative of the 
                detector type and energy range. We exclusively relied on publicly available
                data provided by the collaborations. The data is also available in the
                \code{gammapy-data} repository.
    }
        \label{fig:irfs}
\end{figure*}

The \code{gammapy.irf} sub-package contains all classes and functionalities
to handle IRFs in a variety of functional forms.
Usually, \irfs store instrument properties in the form of multidimensional
tables, with quantities expressed in terms of energy (true or reconstructed),
off-axis angles or cartesian detector coordinates. The main quantities stored in
the common \gammaray \irfs are the effective area, energy dispersion,
PSF and background rate. The \code{gammapy.irf}
sub-package can open and access specific \irf extensions,
interpolate and evaluate the quantities of interest on both energy and spatial
axes, convert their format or units, plot or write them into
output files. In the following, we list the main classes of the
sub-package:

\subsubsection{Effective area}
\gammapy provides the class \code{EffectiveAreaTable2D} to
manage the effective area, which is usually defined in terms of true energy and offset angle.
The class functionalities offer the possibility to read from files or to create
it from scratch. The \code{EffectiveAreaTable2D} class can also convert, interpolate,
write, and evaluate the effective area for a given energy and offset angle, or
even plot the multidimensional effective area table.

\subsubsection{Point spread function}
\gammapy allows users to treat different kinds of PSFs,
in particular, parametric multidimensional Gaussian distributions (\code{EnergyDependentMultiGaussPSF})
or King profile functions (\code{PSFKing}). The \code{EnergyDependentMultiGaussPSF}
class is able to handle up to three Gaussians, defined in terms of amplitudes and sigma given for each true energy
and offset angle bin. The general \code{ParametricPSF} class allows users to create a
custom PSF with a parametric representation different from Gaussian(s) or King profile(s).
The generic \code{PSF3D} class stores a radial symmetric profile of a
PSF to represent nonparametric shapes, again depending on true energy
and offset from the pointing position.

To handle the change of the PSF with the observational offset during the analysis 
the \code{PSFMap} class is used. It stores the radial profile of the PSF
depending on the true energy and position on the sky. During the modeling
step in the analysis, the PSF profile for each model component is 
looked up at its current position and converted into a 3D convolution kernel
which is used for the prediction of counts from that model component.

\subsubsection{Energy dispersion}
For \iacts, the energy resolution and bias, sometimes called energy dispersion,
is typically parameterized in terms of the so-called
migration parameter ($\mu$, see section \label{sec:gammaray-data-analysis}).
The distribution of the migration parameter is given at each offset angle and
true energy. The main subclasses are the \code{EnergyDispersion2D} which is
designed to handle the raw instrument description, and the \code{EDispKernelMap},
which contains an energy dispersion matrix per sky position, that is, a 4D
sky map where each position is associated with an energy dispersion matrix.
The energy dispersion matrix is a representation of the energy resolution
as a function of the true energy only and implemented in \gammapy
by the subclass \code{EDispKernel}.

\subsubsection{Instrumental background}
The instrumental background rate can be represented as either a 2D
data structure named \code{Background2D} or a 3D one named \code{Background3D}.
The background rate is stored as a differential count rate, normalized per solid angle
and energy interval at different reconstructed energies and position in the FoV.
In the \code{Background2D} case, the background is expected to follow a radially symmetric shape
and changes only with the offset angle from FoV center.
In the \code{Background3D} case, the background is allowed to vary with 
longitude and latitude of a tangential FoV coordinates system.

Some example IRFs read from public data files and plotted with \gammapy are shown 
in Figure~\ref{fig:irfs}. More information on \code{gammapy.irf} can be found 
in the online documentation\footnote{\url{https://docs.gammapy.org/1.0/user-guide/irf/index.html\#irf}}.

\subsection{gammapy.maps}
\label{ssec:gammapy-maps}
The \code{gammapy.maps} sub-package provides classes that represent data
structures associated with a set of coordinates or a region on a sphere. In
addition it allows one to handle an arbitrary number of nonspatial data
dimensions, such as time or energy. It is organized around three types of
structures: geometries, sky maps and map axes, which inherit from the base
classes \code{Geom}, \code{Map} and \code{MapAxis} respectively.

The geometry object defines the pixelization scheme and map boundaries. It also
provides methods to transform between sky and pixel coordinates. Maps consist
of a geometry instance defining the coordinate system together with a
\numpy array containing the associated data. All map classes support a basic
set of arithmetic and boolean operations with  unit support, up and downsampling
along extra axes, interpolation, resampling of extra axes, interactive visualization
in notebooks and interpolation onto different geometries.

The \code{MapAxis} class provides a uniform application programming interface
(API) for axes representing
bins on any physical quantity, such as energy or angular offset.
Map axes can have physical units attached to them, as well as define
nonlinearly spaced bins. The special case of time is covered by the
dedicated \code{TimeMapAxis}, which allows time bins to be noncontiguous,
as it is often the case with observational times. The generic
class \code{LabelMapAxis} allows the creation of axes for non-numeric entries.

To handle the spatial dimension the sub-package exposes a uniform API for
the FITS World Coordinate System (WCS), the HEALPix pixelization and
region-based data structure (see Figure~\ref{fig*:minted:gp_maps}).
This allows users to perform the same higher level operations on maps
independent of the underlying pixelization scheme. The \code{gammapy.maps}
package is also used by external packages such as \enquote{FermiPy}~\citep{Wood2017}.

\begin{figure}
        \small
        \import{code-examples/generated/}{gp_maps}

        \caption{
        Using \code{gammapy.maps} to create a WCS, a HEALPix and a region
                based data structures. The initialization parameters include
        consistently the positions of the center of the map, the pixel
        size, the extend of the map as well as the energy axis definition.
        The energy minimum and maximum values for the creation of the
        \code{MapAxis} object can be defined as strings also specifying the
        unit. Region definitions can be passed as strings following
        the DS9 region specifications \url{http://ds9.si.edu/doc/ref/region.html}. The output
                of the code example is shown in Figure~\ref{fig:code_example_gp_maps}.
        }
    \label{fig*:minted:gp_maps}
\end{figure}

\subsubsection{WCS maps}
The FITS WCS pixelization supports a number of different projections to
represent celestial spherical coordinates in a regular rectangular grid.
\gammapy provides full support to data structures using this pixelization
scheme. For details see ~\cite{Calabretta2002}. This pixelization
is typically used for smaller regions of interests, such as pointed
observations and is represented by a combination of the
\code{WcsGeom} and \code{WcsNDMap} class.

\subsubsection{HEALPix maps}
This pixelization scheme ~\citep{Calabretta2002, Gorski2005} provides a
subdivision of a sphere in which each pixel covers the same surface area as
every other pixel. As a consequence, however, pixel shapes are no longer
rectangular, or regular.
This pixelization is typically used for all-sky data, such as data
from the \hawc or \fermi observatory. \gammapy natively supports
the multiscale definition of the HEALPix pixelization and thus
allows for easy upsampling and downsampling of the data. In addition to
the all-sky map, \gammapy also supports a local HEALPix
pixelization where the size of the map is constrained to a given
radius.
For local neighborhood operations, such as convolution, \gammapy relies
on projecting the HEALPix data to a local tangential WCS grid.
This data structure is represented by the \code{HpxGeom} and \code{HpxNDMap}
classes.

\subsubsection{Region maps}
In this case, instead of a fine spatial grid
dividing a rectangular sky region, the spatial dimension is reduced to a single
bin with an arbitrary shape, describing a region in the sky with that same
shape. Region maps are typically used together with a nonspatial dimension, for
example an energy axis, to represent how a quantity varies in that dimension
inside the corresponding region. To avoid the complexity of handling
spherical geometry for regions, the regions are projected onto the local
tangential plane using a WCS transform. This approach follows Astropy's \enquote{Regions}
package \citep{AstropyRegions2022}, which is both used as an API to define regions
for users as well as handling the underlying geometric operations. Region based
maps are represented by the \code{RegionGeom} and \code{RegionNDMap} classes.
More information on the \code{gammapy.maps} sub-module can be found 
in the online documentation\footnote{\url{https://docs.gammapy.org/1.0/user-guide/maps/index.html\#maps}}.

\subsection{gammapy.datasets}
\label{ssec:gammapy-datasets}
\begin{figure}
        \small
        \import{code-examples/generated/}{gp_datasets}
        \caption{
        Using \code{gammapy.datasets} to read existing reduced binned datasets.
        After the different datasets are read from disk they are collected into a
        common \code{Datasets} container. All dataset types have an associated
        name attribute to allow a later access by name in the code. The
        environment variable \code{\$GAMMAPY\_DATA} is automatically resolved
        by \gammapy. The output
                of the code example is shown in Figure~\ref{fig:code_example_gp_datasets}.
    }
        \label{fig*:minted:gp_datasets}
\end{figure}
The \code{gammapy.datasets} sub-package contains classes to bundle
together binned data along with the associated models and likelihood function, which
provides an interface to the \code{Fit} class (Sec \ref{sssec:fit}) for
modeling and fitting purposes. Depending upon the type of analysis and the
associated statistic, different types of Datasets are supported. The \code{MapDataset} is
used for combined spectral and morphological (3D) fitting, while spectral 
fitting only can be performed using the \code{SpectrumDataset}.
While the default fit statistics for both of these classes is the \emph{Cash}~\citep{Cash}
statistic, there are other classes which support
analyses where the background is measured from control regions, so called \enquote{off} observations.
Those require the use of a different fit statistics, which takes into account the
uncertainty of the background measurement. This case is covered by the \code{MapDatasetOnOff}
and \code{SpectrumDatasetOnOff} classes, which use the \emph{WStat}~\citep{WStat} statistic.

Fitting of precomputed flux points is enabled through \code{FluxPointsDataset},
using \emph{$\chi^2$} statistics. Multiple datasets of same or different types can be
bundled together in \code{Datasets} (see e.g., Figure \ref{fig*:minted:gp_datasets}),
where the likelihood from each constituent member is added, thus facilitating
joint fitting across different observations, and even different instruments
across different wavelengths. Datasets also provide functionalities for
manipulating reduced data, for instance stacking, sub-grouping, plotting. Users can
also create their customized datasets for implementing modified likelihood
methods. We also refer to our online documentation for more details on \code{gammapy.datasets}
\footnote{\url{https://docs.gammapy.org/1.0/user-guide/datasets/index.html\#datasets}}.

\subsection{gammapy.makers}
\label{ssec:gammapy-makers}
\begin{figure}
        \small
        \import{code-examples/generated/}{gp_makers}
        \caption{
        Using \code{gammapy.makers} to reduce DL3 level data into a
                \code{MapDataset}. All \code{Maker} classes represent 
                a step in the data reduction process. They take
        the configuration on initialization of the class. They 
                also consistently define \code{.run()} methods, which take
                a dataset object and optionally an \code{Observation} 
                object. In this way, \code{Maker} classes can be chained
                to define more complex data reduction pipelines. The output
                of the code example is shown in Figure~\ref{fig:code_example_gp_makers}.
    }
        \label{fig*:minted:gp_makers}
\end{figure}
The \code{gammapy.makers} sub-package contains the various classes and functions required
to process and prepare \gammaray data from the DL3 to the DL4.
The end product of the data reduction process is a set of binned counts,
background exposure, psf and energy dispersion maps at the DL4 level, bundled
into a \code{MapDataset} object.
The \code{MapDatasetMaker} and \code{SpectrumDatasetMaker} are
responsible for this task for three- and 1D analyses, respectively (see Figure~\ref{fig*:minted:gp_makers}).

The correction of background models from the data themselves is supported 
by specific \code{Maker} classes such as the \code{FoVBackgroundMaker} or the
\code{ReflectedRegionsBackgroundMaker}. The former is used to estimate the
normalization of the background model from the data themselves, while the
latter is used to estimate the background from regions reflected from the
pointing position.

Finally, to limit other sources of systematic uncertainties, a data validity
domain is determined by the \code{SafeMaskMaker}. It can be used to limit the
extent of the FoV used, or to limit the energy range to a domain
where the energy reconstruction bias is below a given threshold.

More detailed information on the \code{gammapy.makers} sub-package is available online\footnote{\url{https://docs.gammapy.org/1.0/user-guide/makers/index.html\#makers}}.

\subsection{gammapy.stats}
\label{ssec:gammapy-stats}
The \code{gammapy.stats} subpackage contains the fit statistics and the associated
statistical estimators commonly adopted in \gammaray astronomy. In
general, \gammaray observations count Poisson-distributed events at various sky
positions and contain both signal and background events.
To estimate the number of signal events in the observation one typically uses
Poisson maximum likelihood estimation (MLE). In practice this is done
by minimizing a fit statistic defined by $-2 \time\cdot \log{\mathcal{L}}$,
where $\mathcal{L}$ is the likelihood function used. \gammapy uses the convention
of a factor of 2 in front, such that a difference in log-likelihood will
approach a $\chi^2$ distribution in the statistial limit.

When the expected number of background events is known, the statistic function 
is the so called \emph{Cash} statistic ~\citep{Cash}. It is used by datasets using background
templates such as the
\code{MapDataset}. When the number of background events is unknown, and an \enquote{off}
measurement where only background events are expected is used, the statistic
function is \emph{WStat}. It is a profile log-likelihood statistic where the background
counts are marginalized parameters. It is used by datasets containing \enquote{off}
counts measurements such as the \code{SpectrumDatasetOnOff}, used for classical
spectral analysis.

To perform simple statistical estimations on counts measurements,
\code{CountsStatistic} classes encapsulate the aforementioned statistic functions to
measure excess counts and estimate the associated statistical significance,
errors and upper limits. They perform maximum likelihood ratio tests to
estimate significance (the square root of the statistic difference) and compute
likelihood profiles to measure errors and upper limits. The code example
\ref{fig*:minted:gp_stats} shows how to compute the Li \& Ma
significance~\citep{LiMa} of a set of measurements.
Our online documentation provides more information on \code{gammapy.stats}
\footnote{\url{https://docs.gammapy.org/1.0/user-guide/stats/index.html\#stats}}. 

\begin{figure}
        \small
        \import{code-examples/generated/}{gp_stats}
        \caption{
        Using \code{gammapy.stats} to compute statistical quantities
        such as excess, significance and asymmetric errors
        from counts based data. The data array such as \code{counts}, \code{counts\_off}
                and the background efficiency ratio \code{alpha} are passed on initialization
        of the \code{WStatCountsStatistic} class. The derived quantities
        are then computed dynamically from the corresponding class
        attributes such as \code{stat.n\_sig} for the excess
                and \code{stat.sqrt\_ts} for the significance.
                The output of the code example is shown in Figure~\ref{fig:code_example_gp_stats}.
    }
        \label{fig*:minted:gp_stats}
\end{figure}

\subsection{gammapy.modeling}
\label{ssec:gammapy-modeling}
\code{gammapy.modeling} contains all the functionality related to modeling and fitting
data. This includes spectral, spatial and temporal model classes, as well as
the fit and parameter API.

\subsubsection{Models}
\label{sssec:models}
Source models in \gammapy (Eq.~\ref{eq:general_source_model}) are 4D 
analytical models which support two spatial dimensions defined by the sky coordinates
$\ell, b$, an energy dimension $E$, and a time dimension $t$. To simplify the definition of the
models, \gammapy uses a factorized representation of the total source
model:

\begin{equation}
    \phi(\ell, b, E, t) = F(E) \cdot G(\ell, b, E) \cdot H(t, E).
    \label{eq:source_model_dependency}
\end{equation}

The spectral component $F(E)$, described by the \code{SpectralModel} class, always
includes an amplitude parameter to adjust the total flux of the model.
The spatial component $G(\ell, b, E)$, described by the \code{SpatialModel} class,
also depends on energy, in order to consider energy-dependent sources morphology.
Finally, the temporal component $H(t, E)$, described by the \code{TemporalModel}
class, also supports an energy dependency in order to consider spectral variations
of the model with time.

The models follow a naming scheme which contains the category as a suffix to
the class name. The spectral models include a special class of normed models,
named using the \code{NormSpectralModel} suffix.
These spectral models feature a dimension-less \enquote{norm} parameter
instead of an amplitude parameter with physical units. They
can be used as an energy-dependent multiplicative correction
factor to another spectral model. They are typically used for
adjusting template-based models, or, for example, to take into account
the absorption effect on \gammaray spectra caused by the extra-galactic background
light (EBL) (\code{EBLAbsorptionNormSpectralModel}). \gammapy supports a variety
of EBL absorption models, such as those from \cite{Franceschini2008, Franceschini2017}, \cite{Finke2010},
and \cite{Dominguez2011}.

The analytical spatial models are all normalized such that they integrate to
unity over the entire sky. The template spatial models may not, so in that special
case they have to be combined with a \code{NormSpectralModel}.

The \code{SkyModel} class represents the factorized model in Eq.~\ref{eq:source_model_dependency}
(the spatial and temporal components being optional).
A \code{SkyModel} object can represent the sum of several emission components:
either, for example, from multiple sources and from a diffuse emission, or from several spectral
components within the same source. To handle a list of multiple \code{SkyModel} objects, \gammapy
implements a \code{Models} class.

The model gallery\footnote{\url{https://docs.gammapy.org/1.0/user-guide/model-gallery/index.html}}
 provides a visual overview of the available models in
\gammapy. Most of the analytic models commonly used in \gammaray astronomy are
built-in. We also offer a wrapper to radiative models implemented in the Naima
package~\citep{naima}. The modeling framework can be easily extended with
user-defined models. For example, the radiative models of jetted Active Galactic Nuclei (AGN)
implemented in \agnpy~\citep{agnpy2022}, can be wrapped into 
\gammapy~\citep[see Section 3.5 of ][]{2022A&A...660A..18N}. We provide 
more examples of user-defined models, such as a parametric model for energy dependent morphology,
in the online documentation\footnote{\url{https://docs.gammapy.org/1.0/tutorials/api/models.html\#implementing-a-custom-model}}.

\begin{figure}
        \small
        \import{code-examples/generated/}{gp_models}
        \caption{Using \code{gammapy.modeling.models} to define a source model with a
    spectral, spatial and temporal component. For convenience the model
    parameters can be defined as strings with attached units. The spatial model
    takes an additional \code{frame} parameter which allow users to define
    the coordinate frame of the position of the model. The output
        of the code example is shown in Figure~\ref{fig:code_example_gp_models}.
    }
        \label{fig*:minted:gp_models}
\end{figure}

\subsubsection{Fit}
\label{sssec:fit}

The \code{Fit} class provides methods to optimize (\enquote{fit}), model parameters and estimate
their errors and correlations. It interfaces with a \code{Datasets} object, which
in turn is connected to a \code{Models} object containing the model parameters bundled into a
\code{Parameters} object. Models can be unique for a given dataset, or contribute to
multiple datasets, allowing one to perform a joint fit to
multiple IACT datasets, or to jointly fit IACT and \fermi datasets. Many
examples are given in the tutorials.

The \code{Fit} class provides a uniform interface to multiple fitting backends:
\begin{itemize}
        \setlength\itemsep{1em}
        \item \iminuit~\citep{iminuit}
        \item \code{scipy.optimize}~\citep{2020SciPy-NMeth}
        \item \sherpa~\citep{sherpa-2011, sherpa-2001}
\end{itemize}

We note that, for now, covariance matrix and errors are computed only for the fitting with 
\iminuit. However, depending on
the problem other optimizers can perform better, so sometimes it can be useful
to run a pre-fit with alternative optimization methods. In the future, we plan to
extend the supported fitting backends, including for example solutions based on Markov chain Monte Carlo methods.
\footnote{a prototype is available in gammapy-recipes,
        \url{https://gammapy.github.io/gammapy-recipes/_build/html/notebooks/mcmc-sampling-emcee/mcmc_sampling.html}
}

\subsection{gammapy.estimators}
\label{ssec:gammapy-estimators}
By fitting parametric models to the data, the total \gammaray
flux and its overall temporal, spectral and morphological components can be constrained.
In many cases though, it is useful to make a more detailed follow-up analysis by measuring the
flux in smaller spectral, temporal or spatial bins. This
possibly reveals more detailed emission features, which
are relevant for studying correlation with counterpart emissions.

The \code{gammapy.estimators} sub-module features methods to compute flux
points, light curves, flux maps and flux profiles from data.
The basic method for all these measurements is equivalent.
The initial fine bins of \code{MapDataset} are grouped into
larger bins. A multiplicative correction factor (the norm)
is applied to the best fit reference spectral
model and is fitted in the restricted data range, defined by the 
bin group only.

In addition to the best-fit flux norm, all estimators compute
quantities corresponding to this flux. This includes:
the predicted number of total, signal and background
counts per flux bin; the total fit statistics
of the best fit model (for signal and background); the fit statistics of the
null hypothesis (background only); and the difference between both,
the so-called test statistic value (TS).
From this TS value, a significance of the measured signal and associated flux
can be derived.

Optionally, the estimators can also compute more advanced quantities
such as asymmetric flux errors, flux upper limits
and 1D profiles of the fit statistic,
which show how the likelihood functions varies with
the flux norm parameter around the fit minimum.
This information is useful in inspecting the quality
of a fit, for which a parabolic
shape of the profile is asymptomatically expected at the best fit
values.

The base class of all algorithms is the \code{Estimator}  class.
The result of the flux point estimation are either stored in a
\code{FluxMaps} or \code{FluxPoints} object. Both objects
are based on an internal representation of the flux which is
independent of the Spectral Energy Distribution (SED) type. The flux is represented
by a reference spectral model and an array of
normalization values given in energy, time and spatial bins,
which factorizes the deviation of the flux in a given
bin from the reference spectral model. This allows
users to conveniently transform between different
SED types. Table~\ref{tab:sed_types} shows an
overview and definitions of the supported SED types.
The actual flux values for each SED type are obtained
by multiplication of the {norm} with the reference flux.

\begin{table*}
    \begin{center}
        \begin{tabular}{lll}
         \hline
         Type & Description & Unit Equivalency \\
         \hline
         dnde & Differential flux at a given energy & \si{TeV^{-1}.cm^{-2}.s^{-1}} \\
         e2dnde & Differential flux at a given energy  & \si{TeV.cm^{-2}.s^{-1}} \\
         flux & Integrated flux in a given energy range & \si{cm^{-2}.s^{-1}} \\
         eflux & Integrated energy flux in a given energy range & \si{erg.cm^{-2}.s^{-1}}\\
         \hline
        \end{tabular}
    \end{center}
    \caption{Definition of the different SED types supported in \gammapy.}
    \label{tab:sed_types}
\end{table*}

Both result objects support the possibility to serialize
the data into multiple formats. This includes the
\gadf SED format \footnote{\url{https://gamma-astro-data-formats.readthedocs.io/en/latest/spectra/flux_points/index.html}},
FITS-based nd sky maps and other formats compatible with Astropy's \code{Table} and
\code{BinnedTimeSeries} data structures. This allows
users to further analyze the results with Astropy, for example using
standard algorithms for time analysis, such as
the Lomb-Scargle periodogram or the Bayesian
blocks. So far, \gammapy does not support unfolding of \gammaray spectra.
Methods for this will be implemented in future versions of \gammapy.

The code example shown in Figure~\ref{fig*:minted:gp_estimators} shows how to use
the \code{TSMapEstimator} objects with a given input \code{MapDataset}.
In addition to the model, it allows the energy
bins of the resulting flux and TS maps to be specified.

More details on the \code{gammapy.estimators} sub-module are available online \footnote{\url{https://docs.gammapy.org/1.0/user-guide/estimators.html\#estimators}}.

\begin{figure}
        \small
        \import{code-examples/generated/}{gp_estimators}
        \caption{Using the \code{TSMapEstimator} object from \code{gammapy.estimators} to compute a
         flux, flux upper limits and TS map. The additional parameters \code{n\_sigma}
        and \code{n\_sigma\_ul} define the confidence levels (in multiples of the normal distribution width)
        of the flux error and flux upper limit maps respectively. The output
                of the code example is shown in Figure~\ref{fig:code_example_gp_estimators}.
    }
    \label{fig*:minted:gp_estimators}
\end{figure}

\subsection{gammapy.analysis}
\label{ssec:gammapy-analysis}
The \code{gammapy.analysis} sub-module provides a high-level interface (HLI) for the most
common use cases identified in \gammaray analyses. The included classes and methods
 can be used in Python scripts, notebooks or as commands within \texttt{IPython}
sessions. The HLI can also be used to automatize
workflows driven by parameters declared in a configuration file in YAML format.
In this way, a full analysis can be executed via a single command line taking the
configuration file as input.

The \code{Analysis} class has the responsibility for orchestrating the workflow
defined in the configuration \code{AnalysisConfig} objects and triggering the execution of
the \code{AnalysisStep} classes that define the identified common use cases. These
steps include the following: observations selection with the \code{DataStore},  data
reduction, excess map computation, model fitting, flux points estimation, and
light curves production.

\subsection{gammapy.visualization}
\label{ssec:gammapy-visualization}
The \code{gammapy.visualization} sub-package contains helper functions
for plotting and visualizing analysis results and \gammapy~data structures.
This includes, for example, the visualization of reflected background regions across
multiple observations, or plotting large parameter correlation matrices of
\gammapy models. It also includes a helper class to split
wide field Galactic survey images across multiple panels to fit a standard
paper size.

The sub-package also provides \texttt{matplotlib} implementations of specific
colormaps. Those colormaps have been historically used by larger collaborations
in the VHE domain (such as \milagro or \hess) as \enquote{trademark}
colormaps. While we explicitly discourage the use of those colormaps for publication
of new results, because they do not follow modern visualization
standards, such as linear brightness gradients and accessibility
for visually impaired people, we still consider the colormaps
useful for reproducibility of past results.

\subsection{gammapy.astro}
\label{ssec:gammapy-astro}
The \code{gammapy.astro} sub-package contains utility functions for studying physical
scenarios in high-energy astrophysics. The \code{gammapy.astro.darkmatter} module
computes the so called J-factors and the associated \gammaray spectra expected
from annihilation of dark matter in different channels, according to the recipe
described in \cite{2011JCAP...03..051C}.

In the \code{gammapy.astro.source} sub-module, dedicated classes exist for modeling
galactic \gammaray sources according to simplified physical models, for example supernova remnant (SNR) evolution
models \citep{1950RSPSA.201..159T, 1999ApJS..120..299T}, evolution of pulsar wind nebulae (PWNe) during the
free expansion phase \citep{2006ARA&A..44...17G}, or computation
of physical parameters of a pulsar using a simplified dipole spin-down model.

In the \code{gammapy.astro.population} sub-module there are dedicated tools
for simulating synthetic populations based on physical models derived from
observational or theoretical considerations for different classes of Galactic
very high-energy \gammaray emitters: PWNe, SNRs \cite{1998ApJ...504..761C},
pulsars \cite{2006ApJ...643..332F, 2006MNRAS.372..777L, 2004A&A...422..545Y}
and \gammaray binaries.

While the present list of use cases is rather preliminary, this can be enriched
with time by users and/or developers according to future needs.

\subsection{gammapy.catalog}
\label{ssec:gammapy-catalog}
Comprehensive source catalogs are increasingly being provided by many high-energy
astrophysics experiments. The \code{gammapy.catalog} sub-packages
provides a convenient access to the most important \gammaray catalogs.
Catalogs are represented by the \code{SourceCatalog} object, which
contains the actual catalog as an Astropy \code{Table} object.
Objects in the catalog can be accessed by row index, name of the
object or any association or alias name listed in the catalog.

Sources are represented in \gammapy by the \code{SourceCatalogObject}
class, which has the responsibility to translate the information
contained in the catalog to \gammapy objects. This includes
the spatial and spectral models of the source, flux points and
light curves (if available) for each individual object. 
Figure ~\ref{fig*:minted:gp_catalogs}
show how to load a given catalog and access these information for a selected source.
The required catalogs are supplied in the \code{GAMMAPY\_DATA} repository, such that 
user do not have to download them separately.
The overview of currently supported catalogs, the corresponding
\gammapy classes and references are shown in Table~\ref{tab:catalogs}.
Newly released relevant catalogs will be added in future.

\begin{table*}[ht!]
    \begin{center}
        \begin{tabular}{llll}
         \hline
         Class Name & Shortcut & Description & Reference\\
         \hline
         \code{SourceCatalog3FGL} & \code{"3fgl"} & 3\textsuperscript{rd} catalog of \fermi sources & \cite{3FGL} \\
         \code{SourceCatalog4FGL} & \code{"4fgl"} & 4\textsuperscript{th} catalog of \fermi  sources & \cite{4FGL} \\
         \code{SourceCatalog2FHL} & \code{"2fhl"} & 2\textsuperscript{nd} catalog high-energy \fermi  sources & \cite{2FHL} \\
         \code{SourceCatalog3FHL} & \code{"3fhl"} & 3\textsuperscript{rd} catalog high-energy \fermi  sources & \cite{3FHL} \\
         \code{SourceCatalog2HWC} & \code{"2hwc"} & 2\textsuperscript{nd} catalog of \hawc sources & \cite{2HWC} \\
         \code{SourceCatalog3HWC} & \code{"3hwc"} & 3\textsuperscript{rd} catalog of \hawc sources & \cite{3HWC} \\
         \code{SourceCatalogHGPS} & \code{"hgps"} & \hess Galactic Plane Survey catalog & \cite{HGPS} \\
         \code{SourceCatalogGammaCat} & \code{"gammacat"} & Open source data collection & \cite{gamma-cat} \\
         \hline
         \end{tabular}
    \end{center}
    \caption{Overview of supported catalogs in \code{gammapy.catalog}.}
    \label{tab:catalogs}
\end{table*}

\begin{figure}
        \small
        \import{code-examples/generated/}{gp_catalogs}
        \caption{Using \code{gammapy.catalogs} to access the underlying model, flux points and
                light-curve from the \fermi 4FGL catalog for the blazar PKS 2155-304. The output
                of the code example is shown in Figure~\ref{fig:code_example_gp_catalogs}.
        }
        \label{fig*:minted:gp_catalogs}
\end{figure}

\section{Applications}
\label{sec:applications}
\gammapy is currently used for a variety of analyses by different IACT
experiments and has already been employed in about 65 scientific publications as of 21/03/2023
\footnote{\href{https://ui.adsabs.harvard.edu/search/q=(\%20(citations(doi\%3A\%2210.1051\%2F0004-6361\%2F201834938\%22)\%20OR\%20citations(bibcode\%3A2017ICRC...35..766D))\%20AND\%20year\%3A2014-2023)&sort=date\%20desc\%2C\%20bibcode\%20desc&p_=0}{List on ADS}}.
In this section, we illustrate the capabilities of \gammapy by performing some standard
analysis cases commonly considered in \gammaray astronomy.
Beside reproducing standard methodologies, we illustrate the unique data combination
capabilities of \gammapy by presenting a multi-instrument analysis, which is not possible within any
of the current instrument private software frameworks.
The examples shown are based on the data accessible in the \code{gammapy-data} repository,
and limited by the availability of public data.
We remark that, as long as the data are compliant with the GADF specifications (or its future evolutions),
and hence with Gammapy's data structures, there is no limitation on performing
analyses of data from a given instrument.

\subsection{1D analysis}
\label{ssec:1d-analysis}
\begin{figure*}
        \centering
        \includegraphics[width=1.\textwidth]{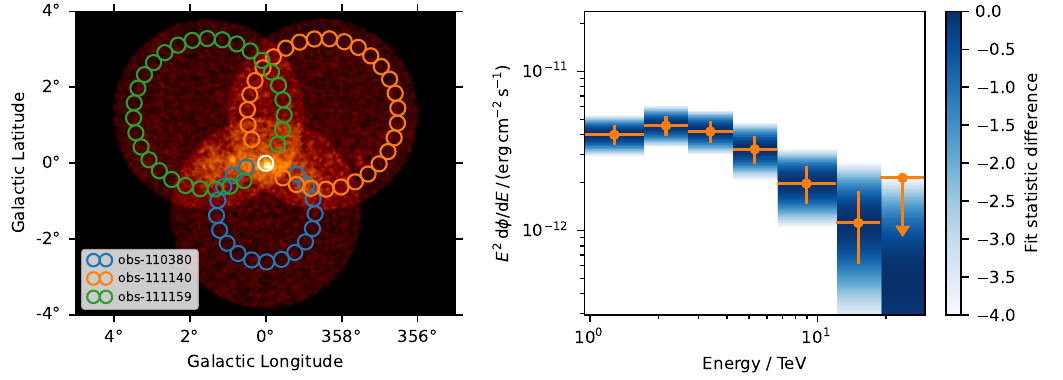}
        \caption{
                Example of a 1D spectral analysis of the Galactic Center for three simulated
                observations from the first CTA data challenge. The left image shows the maps of counts with the signal
                region in white and the reflected background regions for the three different observations
                overlaid in different colors. The right image shows the resulting spectral flux points and
                their corresponding log-likelihood profiles. The flux points are shown in orange, with the
                horizontal bar illustrating the width of the energy bin and the vertical bar the $1~\sigma$
                error. The log-likelihood profiles for each enetgy bin are shown in the background. The colormap
                illustrates the difference of the log-likelihood to the log-likelihood of the best fit value.
                }
        \label{fig:cta_galactic_center}
\end{figure*}

One of the most common analysis cases in \gammaray astronomy is measuring the
spectrum of a source in a given region defined on the sky, in conventional
astronomy also called \emph{aperture photometry}. The spectrum is typically measured
in two steps: first a parametric spectral model is fitted to the data and
secondly flux points are computed in a predefined set of energy bins. The
result of such an analysis performed on three simulated CTA observations is
shown in Figure~\ref{fig:cta_galactic_center}. In this case the spectrum was
measured in a circular aperture centered on the Galactic Center, in
\gammaray~astronomy often called \enquote{on region}. For such analysis the user first
chooses a region of interest and energy binning, both defined by a
\code{RegionGeom}. In a second step, the events and the IRFs are binned
into maps of this geometry, by the \code{SpectrumDatasetMaker}. All the data and
reduced IRFs are bundled into a \code{SpectrumDataset}. To estimate
the expected background in the \enquote{on region} a \enquote{reflected regions} background
method was used~\citep{Berge07}, represented in \gammapy by the
\code{ReflectedRegionsBackgroundMaker} class. The resulting reflected regions are
illustrated for all three observations overlaid on the counts map in Figure~\ref{fig:cta_galactic_center}.
After reduction, the data were modeled using a forward-folding method and assuming
a point source with a power law spectral shape. The model was defined, using
the \code{SkyModel} class with a \code{PowerLawSpectralModel} spectral component only.
This model was then combined with the \code{SpectrumDataset}, which contains the reduced data 
and fitted using the \code{Fit} class. Based on this best-fit model, the final flux points and corresponding
log-likelihood profiles were computed using the \code{FluxPointsEstimator}. The
example takes $<10$ seconds to run on a standard laptop with M1 arm64 CPU.

\subsection{3D analysis}
\label{ssec:3d-analysis}
\begin{figure*}[t]
        \centering
        \includegraphics[width=1.\textwidth]{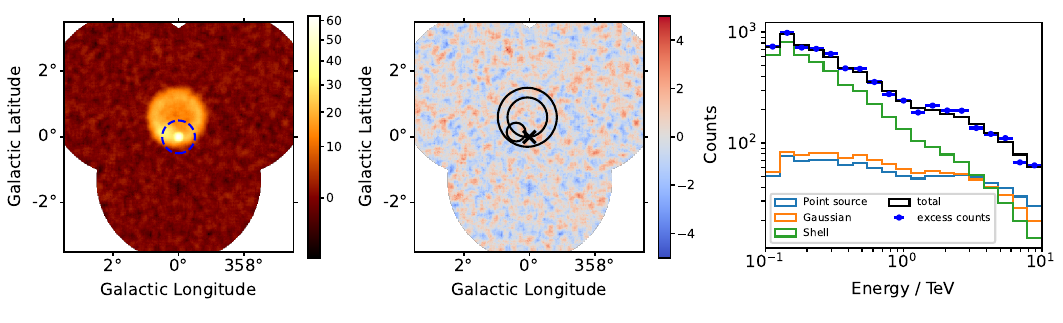}
        \caption{Example of a 3D analysis for simulated sources with point-like, Gaussian
                and shell-like morphologies. The simulation uses \enquote{prod5} \irfs from \cta
                \citep{CtaProd5}.
                The left image shows a significance map (using the \emph{Cash} statistics)
                where the three simulated sources can be seen. The middle figure shows another significance map,
                but this time after
                subtracting the best-fit model for each of the sources, which are displayed in
                black. The right figure shows the contribution of each source model to the
                circular region of radius 0.5\textdegree~drawn in the left image, together with
                the excess counts inside that region. }
        \label{fig:cube_analysis}
\end{figure*}
The 1D analysis approach is a powerful tool to measure the spectrum of an
isolated source. However, more complicated situations require a more careful
treatment. In a FoV containing several overlapping sources, the 1D
approach cannot disentangle the contribution of each source to the total flux in
the selected region. Sources with extended or complex morphology can result in
the measured flux being underestimated, and heavily dependent on the choice of
extraction region.

For such situations, a more complex approach is needed, the so-called 3D
analysis. The three relevant dimensions are the two spatial angular coordinates
and an energy axis. In this framework, a combined spatial and spectral model
(that is, a \code{SkyModel}, see Section~\ref{ssec:gammapy-modeling}) is fitted to the
sky maps that were previously derived from the data reduction step and bundled into a
\code{MapDataset} (see Sections~\ref{ssec:gammapy-makers} and~\ref{ssec:gammapy-datasets}).

A thorough description of the 3D analysis approach and multiple examples that
use \gammapy can be found in~\cite{Mohrmann2019}. Here we present a short
example to highlight some of its advantages.

Starting from the \irfs corresponding to the same three simulated \cta
observations used in Section~\ref{ssec:1d-analysis}, we can create a \code{MapDataset}
via the \code{MapDatasetMaker}. However, we will not use the simulated event lists
provided by \cta but instead, use the method \code{MapDataset.fake()} to simulate
measured counts from the combination of several \code{SkyModel} instances. In this
way, a DL4 dataset can directly be simulated. In particular we simulate:
\begin{enumerate}
    \item a point source located at (l=0\textdegree, b=0\textdegree) with a power law
              spectral shape,
    \item an extended source with Gaussian morphology located at (l=0.4\textdegree,
              b=0.15\textdegree) with $\sigma$=0.2\textdegree~and a log parabola spectral
              shape,
    \item a large shell-like structure centered on (l=0.06\textdegree,
              b=0.6\textdegree) with a radius and width of 0.6\textdegree~and 0.3\textdegree~
              respectively and a power law spectral shape.
\end{enumerate}

The position and sizes of the sources
have been selected so that their contributions overlap. This can be clearly
seen in the significance map shown in the left panel of
Figure~\ref{fig:cube_analysis}. This map was produced with the
\code{ExcessMapEstimator} (see Section~\ref{ssec:gammapy-estimators}) with a
correlation radius of 0.1\textdegree.

We can now fit the same model shapes to the simulated data and retrieve the
best-fit parameters. To check the model agreement, we compute the residual
significance map after removing the contribution from each model. This is done
again via the \code{ExcessMapEstimator}. As can be seen in the middle panel of
Figure~\ref{fig:cube_analysis}, there are no regions above or below $5\,\sigma$,
meaning that the models describe the data sufficiently well.

As the example above shows, the 3D analysis allows the
morphology of the emission to be characterized and fit it together with the spectral properties of
the source.  Among the advantages that this provides is the ability to
disentangle the contribution from overlapping sources to the same spatial
region. To highlight this, we define a circular \code{RegionGeom} of radius
0.5\textdegree~ centered around the position of the point source, which is drawn
in the left panel of Figure~\ref{fig:cube_analysis}. We can now compare the
measured excess counts integrated in that region to the expected relative
contribution from each of the three source models. The result can be seen in the right
panel of Figure~\ref{fig:cube_analysis}. We note that all the models fitted also have a
spectral component, from which flux points can be derived in a similar way as described
in Section~\ref{ssec:1d-analysis}. The whole example takes $<2$ minutes to run on a standard
laptop with M1 arm64 CPU.

\subsection{Temporal analysis}
\label{ssec:temporal-analysis}
\begin{figure*}[t]
    \sidecaption
        \includegraphics[width=0.6666\textwidth]{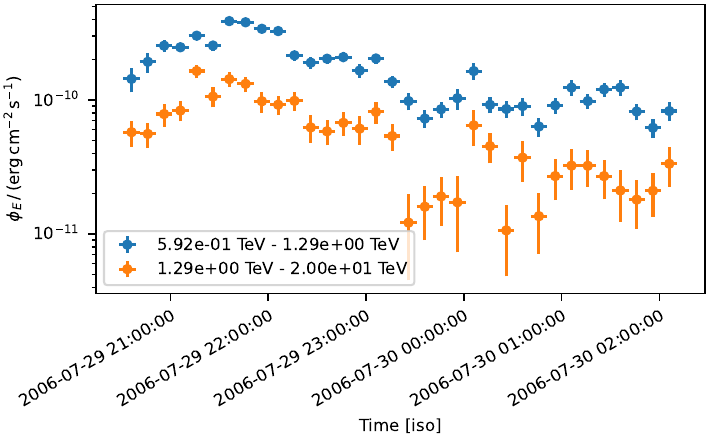}
        \caption{
        Binned PKS~2155-304 light curve in two different energy bands
        as observed by the \hess telescopes in 2006. The colored markers
        show the flux points in the different energy bands:
                the range from (\SI{0.5}{TeV}~to~\SI{1.5}{TeV} is shown in blue, while
                the range from \SI{1.5}{TeV}~to~\SI{20}{TeV}) is shown in orange.
                The horizontal error illustrates the width of the time bin of 10~min. The vertical
        error bars show the associated asymmetrical flux errors. The marker
        is set to the center of the time bin.
    }
    \label{fig:hess_lightcurve_pks}
\end{figure*}

A common use case in many astrophysical scenarios is to study the temporal
variability of a source. The most basic way to do this is to construct a
\enquote{light curve}, which corresponds to measuring the flux of a source
in a set of given time bins. In \gammapy, this
is done by using the \code{LightCurveEstimator} that fits the normalization of a
source in each time (and optionally energy) band per observation, keeping constant
other parameters.
For custom time binning, an observation needs to be split into finer time bins using
the \code{Observation.select\_time} method. Figure~\ref{fig:hess_lightcurve_pks}
shows the light curve of the blazar PKS~2155-304 in different energy bands as
observed by the \hess telescope during an exceptional flare on the night of
July 29 - 30, 2006~\cite{2009A&A...502..749A}. The data are publicly available 
as a part of the HESS-DL3-DR1~\cite{HESS_DR1}. Each observation is first split into 10 min smaller
observations, and spectra extracted for each of these within a 0.11\textdegree~radius
around the source. A \code{PowerLawSpectralModel} is fit to all the datasets, leading
to a reconstructed index of $3.54 \pm 0.02$. With this adjusted spectral model
the \code{LightCurveEstimator} runs directly for two energy bands,
\SI{0.5}{TeV}~to~\SI{1.5}{TeV} and \SI{1.5}{TeV}~to~\SI{20}{TeV} respectively.
The obtained flux points can be analytically modeled using the available or
user-implemented temporal models. Alternatively, instead of  extracting a
light curve, it is also possible to directly fit temporal models to the reduced
datasets. By associating an appropriate \code{SkyModel}, consisting of both temporal
and spectral components, or using custom temporal models with spectroscopic
variability, to each dataset, a joint fit across the datasets will directly
return the best fit temporal and spectral parameters. The light curve data reduction
and computation of flux points takes about $0.5$ minutes to run on a standard laptop with M1 arm64 CPU.

\subsection{Multi-instrument analysis}
\label{ssec:multi-instrument-analysis}
\begin{figure*}[t]
        \sidecaption
        \includegraphics[width=0.666\textwidth]{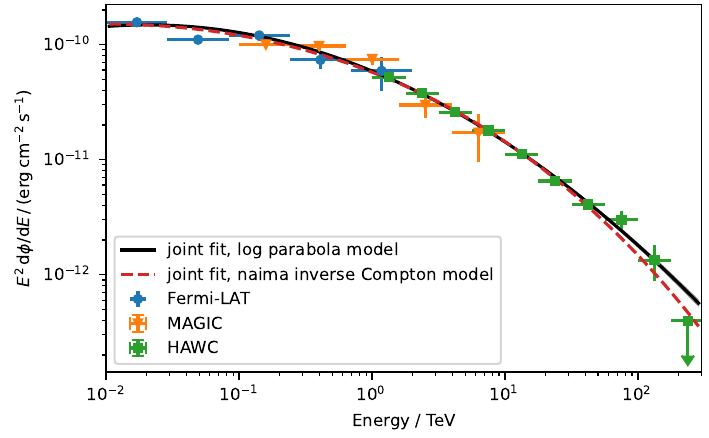}
        \caption{
        Multi-instrument spectral energy distribution (SED) and combined model fit
        of the Crab Nebula. The  colored markers show the flux points computed from
        the data of the different listed instruments. The horizontal error bar
        illustrates the width of the chosen energy band ($E_{Min}, E_{Max}$).
        The marker is set to the log-center energy of the band, that is
        defined by $\sqrt{E_{Min} \cdot E_{Max}}$. The vertical errors bars
        indicate the $1\sigma$ error of the measurement. The downward
        facing arrows indicate the value of $2\sigma$ upper flux limits
        for the given energy range. The black solid line shows the best
        fit model and the transparent band its $1\sigma$ error range.
                The band is too small be visible.
    }
        \label{fig:multi_instrument_analysis}
\end{figure*}
%
In this multi-instrument analysis example we showcase the capabilities of
\gammapy to perform a simultaneous likelihood fit incorporating data from
different instruments and at different levels of reduction. We estimate the
spectrum of the Crab Nebula combining data from the \fermi, \magic and \hawc
instruments.

The \fermi data is introduced at the data level DL4, and directly bundled in a
\code{MapDataset}. They have been prepared using the standard \enquote{fermitools} \citep{Fermitools2019} and
selecting a region of $5^{\circ}\,\times\,4^{\circ}$ around the
position of the Crab Nebula, applying the same selection criteria of the 3FHL
catalog (7 years of data with energy from \SI{10}{GeV} to \SI{2}{TeV},
~\citealt{3FHL}).

The \magic data is included from the data level DL3. They consist of two
observations of $20\,{\rm min}$ each, chosen from the dataset used to estimate
the performance of the upgraded stereo system~\citep{magic_performance} and
already included in~\cite{joint_crab}. The observations were taken at small
zenith angles ($<30^{\circ}$) in wobble mode~\citep{fomin_1994}, with the
source sitting at an offset of $0.4^{\circ}$ from the FoV center. Their energy
range spans \SI{80}{GeV} to \SI{20}{TeV}. The data reduction for the 1D analysis
is applied, and the data are reduced to a \code{SpectrumDataset} before being fitted.

\hawc data are directly provided as flux points (DL5 data level) and are read
via Gammapy's \code{FluxPoints} class. They were estimated in ~\cite{hawc_crab_2019}
with $2.5\,{\rm years}$ of data and span an energy range \SI{300}{GeV} to \SI{300}{TeV}.

Combining the datasets in a \code{Datasets} list, \gammapy automatically generates
a likelihood including three different types of terms, two Poissonian likelihoods
for \fermi's \code{MapDataset} and MAGIC's \code{SpectrumDataset}, and a $\chi^2$
accounting for the \hawc flux points. For \fermi, a 3D forward folding
of the sky model with the IRF is performed, in order to compute the predicted counts
in each sky-coordinate and energy bin. For \magic, a 1D forward-folding
of the spectral model with the \irfs is performed to predict the counts in each estimated energy bin. A log
parabola is fitted over almost five decades in energy \SI{10}{GeV} to \SI{300}{TeV}, taking into account all flux points from all three datasets.

The result of the joint fit is displayed in
Figure~\ref{fig:multi_instrument_analysis}. We remark that the objective of this
exercise is illustrative. We display the flexibility of \gammapy in
simultaneously fitting multi-instrument data even at different levels of
reduction, without aiming to provide a new measurement of the Crab Nebula
spectrum. The spectral fit takes $<10$ seconds to run on a standard laptop with M1 arm64 CPU.


\subsection{Broadband SED modeling}
\label{ssec:broadband-sed-modeling}
By combining \gammapy with astrophysical modeling codes, users can also fit
astrophysical spectral models to \gammaray data. 
There are several Python packages that are able to model
the \gammaray emission, given a physical scenario. Among those
packages are Agnpy~\citep{agnpy2022}, Naima~\citep{naima}, Jetset~\citep{jetset}
and Gamera~\citep{gamera}.
Typically those emission models predict broadband emission from
radio, up to very high-energy \gammarays.
By relying on the multiple dataset types in \gammapy those
data can be combined to constrain such a broadband emission model.
\gammapy provides a built-in \code{NaimaSpectralModel} that allows
users to wrap a given astrophysical emission model from the
Naima package and fit it directly to \gammaray data.

As an example application, we use the same multi-instrument
dataset of the Crab Nebula, described in the previous section,
and we apply an inverse
Compton model computed with Naima and wrapped in the \gammapy models
through the \code{NaimaSpectralModel} class. We describe the gamma-ray emission 
with an inverse Compton scenario, considering a log-parabolic
electron distribution that scatters photons from:
\begin{itemize}
        \item the synchrotron radiation produced by
        the very same electrons
        \item near and far infrared photon fields
        \item and the cosmic microwave background (CMB)
\end{itemize}
We adopt the prescription on the target photon fields provided in the documentation of the \enquote{Naima}
package\footnote{\url{https://naima.readthedocs.io/en/stable/examples.html\#crab-nebula-ssc-model}}.
The best-fit inverse Compton spectrum is represented with a red dashed line in
Figure~\ref{fig:multi_instrument_analysis}. The fit of the astrophysical model
takes $<5$ minutes to run on a standard laptop with M1 arm64 CPU.

More examples for modeling the broadband emission of \gammaray sources, which 
partly involve \gammapy are available in the online documentation of the
Agnpy, Naima, Jetset and Gamera packages\footnote{\url{https://agnpy.readthedocs.io/en/latest/tutorials/ssc\_gammapy\_fit.html}},
\footnote{\url{https://naima.readthedocs.io/en/stable/tutorial.html}}, 
\footnote{\url{https://jetset.readthedocs.io/en/stable/user\_guide/documentation\_notebooks/gammapy\_plugin/gammapy\_plugin.html}},
\footnote{\url{http://libgamera.github.io/GAMERA/docs/fitting\_data.html}}.

\subsection{Surveys, catalogs, and population studies}
\label{ssec:surveys-catalogs-and-population-studies}
As a last application example we describe the use of \gammapy for large scale
analyses such as \gammaray surveys, catalogs and population studies.
Early versions of \gammapy were developed in parallel to the preparation of
the \hess Galactic plane survey catalog~\citep[HGPS, ][]{2018A&A...612A...1H} and
the associated PWN and SNR populations studies~\citep{2018A&A...612A...2H,2018A&A...612A...3H}. 

The increase in sensitivity and angular resolution provided by the new generation of
instruments scales up the number of detectable sources and the complexity of 
models needed to represent them accurately. As an example, if we compare the
results of the HGPS to the expectations from the \cta Galactic Plane survey
simulations, we jump from 78 sources detected by \hess to about 500 detectable by
CTA~\citep{Remy2021}. This large increase in the amount of data to analyze
and increase in complexity of modeling scenarios, requires the high-level
analysis software to be both scalabale as well as performant. 

In short, the production of catalogs from \gammaray surveys can be divided in
four main steps: (a) data reduction, (b) object detection, (c) model fitting and model
selection and (d) associations and classification. All steps can either be done directly
with \gammapy or by relying on the seamless integration of \gammapy with the
scientific Python ecosystem. This allows one to rely on third party functionality
wherever needed. A simplified catalog analysis based on \gammapy typically includes:

\begin{itemize}
        \item[a] The \iacts data reduction step is done in the same way described in the 
        previous sections but scaled up to a few thousand observations. This step can be
        trivially parallelized by deploying the \gammapy package on a cluster.
        \item[b] The object detection step typically consists in finding local maxima
        in the significance or TS maps, computed by the \code{ExcessMapEstimator} or
        \code{TSMapEstimator} respectively. For this \gammapy provides a simple \code{find\_peaks}
        method. For more advanced methods users can rely on third party packages such as
        \enquote{Scikit-image}~\citep{scikit-image}. This packages provide for example general
        \enquote{blob detection} algorithms and image segmentation methods such as hysteresis thresholding
        or the watershed transform.     
        \item[c] During the modeling step each object is alternatively fitted with different
        models in order to determine their optimal parameters, and the best-candidate model. The
        subpackage \code{gammapy.modeling.models} offers a large variety of choices for spatial
        and spectral models as well as the possibility to add custom models. For the model selection
        \gammapy provides statisticl helper methods to perform likelihood ratio tests.
        But users can also rely on third party packages such as \enquote{Scikit-learn}~\citep{scikit-learn}
        to compute quantities such as the Akaike information criterion (AIC) or the Bayesian
        information criterion (BIC), which also allow for model selection.
        \item[d] For the association and classification step, which is tightly connected to
        population studies, we can easily compare the fitted models to the set of
        existing \gammaray catalogs available in \code{gammapy.catalog}. Further
        multi-wavelength cross-matches are usually required to characterize the
        sources. This can easily be achieved by relying on coordinate
        handling from Astropy in combination with affiliated packages such as
        \enquote{Astroquery}~\citep{astroquery}. For more advanced source classification
        methods users can again rely for example on Scikit-learn to perform supervised
        or unsupervised clustering.
\end{itemize}

\gammapy has been successfully used for catalog studies performed on simulations
of the future \cta Galactic Plane Survey~\citep{Remy2021}. Besides the scientific
insights they also gave us the opportunity to test the \gammapy software on complex
use cases. This resulted in a number of improvements to the \gammapy package, such as
as improved performance, optimized analysis strategies documented to users, and
identifying the needs for future developments, including solutions for distributed
computing. The catalog studies of the \cta Galactic Plane Survey simulations
also allowed for detailed cross-comparison of the results obtained with the independent
\gammaray analysis package \enquote{ctools} \citep{2016A&A...593A...1K}, with very consistent 
results.

\section{The \gammapy project} \label{sec:gammapy-project}

In this section, we provide an overview of the organization of the \gammapy
project. We briefly describe the main roles and responsibilities within the
team, as well as the technical infrastructure designed to facilitate the
development and maintenance of \gammapy as a high-quality software. We use
common tools and services for software development of Python open-source
projects, code review, testing, package distribution and user support, with a
customized solution for a versioned and thoroughly tested documentation in the form
of user-friendly playable tutorials. This section concludes with an outlook on
the road map for future directions.

\subsection{Organizational structure}
\label{ssec:organizational-structure}

\gammapy is an international open-source project with a broad
developer base and contributions and commitments from multiple groups and
leading institutes in the very high-energy astrophysics
domain\footnote{\url{https://gammapy.org/team.html}}. The main development
road maps are discussed and validated by a \enquote{Coordination Committee}, composed of
representatives of the main contributing institutions and observatories.
This committee is
chaired by a \enquote{Project Manager} and his deputy while two \enquote{Lead Developers} manage
the development strategy and organize technical activities. This
institutionally driven organization, the permanent staff and commitment of
supporting institutes ensure the continuity of the executive teams. A core team
of developers from the contributing institutions is in charge of the regular
development, which benefits from regular contributions of the community at
large.

\subsection{Technical infrastructure}
\label{ssec:technical-infrastructure}

\begin{figure}[t]
        \centering
        \includegraphics[width=0.5\textwidth]{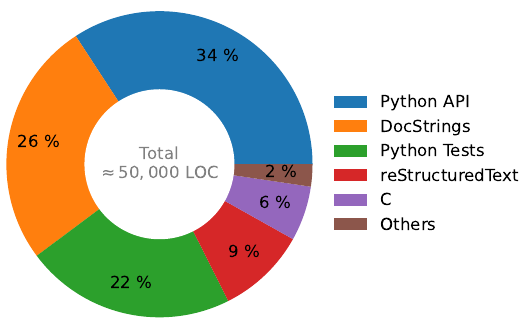}
        \caption{
                Overview of used programming languages and distribution of code across the different file
        categories in the \gammapy code base. The total number of lines is $\approx 50 000$.
    }
        \label{fig:codestats:lang}
\end{figure}

\gammapy follows an open-source and open-contribution development model based on
the cloud repository service \github. A \github organization named
\enquote{gammapy}\footnote{\url{https://github.com/gammapy}} hosts different
repositories related with the project. The software codebase may be found in
the Gammapy repository (see
Figure~\ref{fig:codestats:lang} for code lines statistics). We make extensive
use of the pull request system to discuss and review code contributions.

Several automated tasks are set as \github
actions\footnote{\url{https://github.com/features/actions}}, blocking the
processes and alerting developers when failures occur. This is the case of the
continuous integration workflow, which monitors the execution of the test coverage
suite\footnote{\url{https://pytest.org}} using datasets from the
\enquote{gammapy-data} repository\footnote{\url{https://github.com/gammapy/gammapy-data}}.
Tests scan not only the codebase, but also the
code snippets present in docstrings of the scripts and in the RST documentation
files, as well as in the tutorials provided in the form of Jupyter notebooks.

Other automated tasks, executing in the
\enquote{gammapy-benchmarks}\footnote{\url{https://github.com/gammapy/gammapy-benchmarks}} repository,
are responsible for numerical validation tests and benchmarks monitoring. Also,
tasks related with the release process are partially automated, and every
contribution to the codebase repository triggers the documentation building and
publishing workflow within the
\enquote{gammapy-docs} repository\footnote{\url{https://github.com/gammapy/gammapy-docs}}
(see Sec.~\ref{ssec:software-distribution} and Sec.~\ref{ssec:documentation-and-user-support}).

This small ecosystem of interconnected up-to-date repositories, automated tasks
and alerts, is just a part of a bigger set of \github repositories, where most
of them are related with the project but not necessary for the development of
the software (i.e., project webpage, complementary high-energy astrophysics
object catalogs, coding sprints and weekly developer calls minutes,
contributions to conferences, other digital assets, etc). Finally, third-party
services for code quality metrics are also set and may be found as status
shields in the codebase repository.

\subsection{Software distribution}
\label{ssec:software-distribution}
\gammapy is distributed for Linux, Windows and Mac environments, and installed
in the usual way for Python packages. Each stable release is uploaded to the
Python package index\footnote{\url{https://pypi.org}} and as a binary package
to the \enquote{conda-forge} and \enquote{astropy} Anaconda
repository\footnote{\url{https://anaconda.org/anaconda/repo}} channels. At this
time, \gammapy is also available as a Debian Linux
package\footnote{\url{https://packages.debian.org/sid/python3-gammapy}}. We
recommend installing the software using the \enquote{conda} installation process
with an environment definition file that we provide, so to work within a
virtual isolated environment with additional useful packages and ensure
reproducibility.

\gammapy is indexed in the Astronomy Source Code
Library\footnote{\url{https://ascl.net/1711.014}} and
Zenodo\footnote{\url{https://doi.org/10.5281/zenodo.4701488}} digital libraries for
software. The Zenodo record is synchronized with the codebase \github repository
so that every release triggers the update of the versioned record. In addition,
 \gammapy has been added to the Open-source scientific
Software and Service Repository\footnote{\url{https://projectescape.eu/ossr}} \citep{10.12688/openreseurope.15692.1}
and indexed in the European Open Science Cloud
catalog \footnote{\url{https://eosc-portal.eu}}.

In addition, \gammapy is also listed in the \enquote{SoftWare
Heritage}~\footnote{\url{https://softwareheritage.org}} (SWH) archive~\cite{DiCosmo2020}.
The archive collects, preserves, and shares the source code of publicly available software.
SWH automatically scans open software repositories, such as GitHub, and projects are archived in SWH by the
means of SoftWare Heritage persistent IDentifiers (SWHID), that are guaranteed to remain stable (persistent)
over time. The French open publication archive, HAL~\footnote{\url{https://hal.archives-ouvertes.fr}},
is using the \gammapy SWHIDs to register the releases as scientific
products~\footnote{\url{https://hal.science/hal-03885031v1}} of open science.

\subsection{Documentation and user-support}
\label{ssec:documentation-and-user-support}
\gammapy provides its user community with a tested and versioned up-to-date
online
documentation\footnote{\url{https://docs.gammapy.org}}~\citep{2019ASPC..523..357B}
built with Sphinx\footnote{\url{https://www.sphinx-doc.org}} scanning the
codebase Python scripts, as well as a set of RST files and Jupyter notebooks.
The documentation includes a user guide, a set of executable
tutorials, and a reference to the API automatically extracted from the code and
docstrings. The \gammapy code snippets present in the documentation are tested
in different environments using our continuous integration (CI) workflow based
on \github actions.

The Jupyter notebooks tutorials are generated using the sphinx-gallery
package \citep{sphinx-gallery}.
The resulting web published tutorials also provide links to playground spaces in
\enquote{myBinder}~\citep{project_jupyter-proc-scipy-2018}, where they may be executed
on-line in versioned virtual environments hosted in the myBinder
infrastructure. Users may also play with the tutorials locally in their
laptops. They can download a specific version of the tutorials together with
the associated datasets needed and the specific conda computing environment,
using the \code{gammapy download} command.

We have also set up a solution for users to share recipes
that do not fit in the \gammapy core documentation, but which may be relevant for
specific use cases, in the form of Jupyter notebooks. Contributions happen via pull requests to the
\enquote{gammapy-recipes} \github repository and are merged after a short review. All
notebooks in the repository are tested and published in the \gammapy recipes
webpage\footnote{\url{https://gammapy.github.io/gammapy-recipes}} automatically
using \github actions.

A growing community of users is gathering around the Slack
messaging\footnote{\url{https://gammapy.slack.com}} and \github
discussions\footnote{\url{https://github.com/gammapy/gammapy/discussions}}
support forums, providing valuable feedback on the \gammapy functionalities,
interface and documentation. Other communication channels have been set such as
mailing lists, a Twitter account\footnote{\url{https://twitter.com/gammapyST}},
regular public coding sprint meetings, hands-on sessions within collaborations,
weekly development meetings, etc.

\subsection{Proposals for improving \gammapy}
\label{ssec:pigs}
An important part of \gammapy's development organization is the support
for \enquote{Proposals for improving \gammapy} (PIG). This system is very much
inspired by Python's PEP\footnote{\url{https://peps.python.org/pep-0001/}}
and Astropy's APE \citep{greenfield_perry_2013} system.
PIG are self-contained documents which outline a set of significant
changes to the \gammapy code base. This includes large feature additions,
code and package restructuring and maintenance, as well as changes related
to the organizational structure of the \gammapy project. PIGs can be proposed
by any person in or outside the project and by multiple authors. They
are presented to the \gammapy developer community in a pull request
on \github and then undergo a review phase in which changes and
improvements to the document are proposed and implemented. Once the PIG
document is in a final state it is presented to the \gammapy
coordination committee, which takes the final decision on the
acceptance or rejection of the proposal. Once accepted, the proposed
change are implemented by \gammapy developers in a series of
individual contributions via pull requests. A list of all proposed
PIG documents is available in the \gammapy online documentation
\footnote{\url{https://docs.gammapy.org/dev/development/pigs/index.html}}.

A special category of PIGs are long-term {road maps}. To develop a common
vision for all \gammapy project members on the future of the
project, the main goals regarding planned features, maintenance and
project organization are written up as an overview and presented to the
\gammapy community for discussion. The review and acceptance process
follows the normal PIG guidelines. Typically road maps are written
to outline and agree on a common vision for the next long term
support release of \gammapy.

\subsection{Release cycle, versioning, and long-term support}
\label{ssec:release-cycle}
With the first long term support (LTS) release v1.0, the \gammapy project
enters a new development phase. The development will change from
quick feature-driven development to more stable maintenance
and user support driven development. After v1.0 we foresee
a development cycle with major, minor and bugfix releases;
basically following the development cycle of the Astropy
project. Thus we expect a major LTS release approximately
every two years, minor releases are planned every 6~months,
while bug-fix releases will happen as needed. While
bug-fix releases will not introduce API-breaking changes,
we will work with a deprecation system for minor releases.
API-breaking changes will be announced to users by runtime
warnings first and then implemented in the subsequent
minor release. We consider this approach as a fair
compromise between the interests of users in a stable
package and the interest of developers to improve
and develop \gammapy in future. The development cycle is described
in more detail in PIG 23 \citep{gammapy_pig_23}.

\section{Paper reproducibility}
\label{sec:reproducibility}
One of the most important goals of the \gammapy project is to support open and
reproducible science results. Thus we decided to write this manuscript
openly and publish the Latex source code along with the associated
Python scripts to create the figures
in an open repository~\footnote{\url{https://github.com/gammapy/gammapy-v1.0-paper}}.
This \github repository also documents the history of the creation
and evolution of the manuscript with time. To simplify the reproducibility
of this manuscript including figures and text, we relied on the tool
\enquote{showyourwork}~\citep{Luger2021}. This tool coordinates the building
process and both software and data dependencies, such that the complete
manuscript can be reproduced with a single \code{make} command, after
downloading the source repository. For this we provide
detailed instructions online\footnote{\url{https://github.com/gammapy/gammapy-v1.0-paper/blob/main/README.md}}.
Almost all figures in this manuscript provide a link
to a Python script, that was used to produce it. This means all
example analyses presented in Sec.\ref{sec:applications} link to
actually working Python source code.

\section{Summary and outlook}
\label{sec:summary-and-outlook}
In this paper we have presented the first LTS version of \gammapy.
\gammapy is a Python package for \gammaray astronomy, which relies on the
scientific Python ecosystem, including Numpy, Scipy, and Astropy as
main dependencies. It also holds the status of an Astropy affiliated
package. It supports high-level analysis of astronomical \gammaray
data from intermediate level data formats, such as the FITS based
\gadf. Starting from lists of \gammaray events and corresponding descriptions
of the instrument response, users can reduce and project the data
to WCS, HEALPix, and region-based data structures. The reduced data are bundled
into datasets, which serve as a basis for Poisson maximum likelihood
modeling of the data. For this purpose \gammapy provides a wide selection
of built-in spectral, spatial, and temporal models, as well as a unified
fitting interface with a connection to multiple optimization backends.

With the v1.0 release, the \gammapy project has entered a new development
phase. Future work will not only include maintenance of the v1.0 release,
but also parallel development of new features, improved API, and data
model support. While v1.0 provides all the features required for
standard and advanced astronomical \gammaray data analysis,
we already identified specific improvements to be considered in the
road map for a future v2.0 release. This includes the support for
scalable analyses via distributed computing. This will allow
users to scale an analysis from a few observations to multiple
hundreds of observations, as expected by deep surveys of the CTA
observatory. In addition the high-level interface
of \gammapy is planned to be developed into a fully configurable
API design. This will allow users to define arbitrary complex analysis
scenarios as YAML files and even extend their workflows by user-defined
analysis steps via a registry system. Another important topic will
be to improve the support of handling metadata for data structures
and provenance information to track the history of the data reduction
process from the DL3 to the highest DL5 and DL6 data levels. \gammapy will
also extend its functionalities for time-based analyses,
for example tests for variability in light curves, phase curves peak search, 
as well as improving the interoperability with other timing packages such
as \enquote{Stingray}~\citep{Stingray2019}, Astropy's
time series classes, and \enquote{pint-pulsar}~\citep{Luo2021}
 for high-precision pulsar timing.

Around the core Python package, a large diverse community of
users and contributors has developed. With regular developer meetings,
coding sprints, and in-person user tutorials at relevant conferences
and collaboration meetings, the community has constantly grown.
So far \gammapy has seen ~80 contributors from ten different countries.
With typically ten regular contributors at any given time of the
project, the code base has constantly grown its range of features
and improved its code quality. With \gammapy being officially selected
in 2021 as the base library for the future science tools for CTA
\footnote{\href{https://www.cta-observatory.org/ctao-adopts-the-gammapy-software-package-for-science-analysis/}{CTAO Press Release}},
we expect the community to grow
even further, providing a stable perspective for further usage,
development, and maintenance of the project. In addition to the future use
by the CTA community, \gammapy has already
been used for analysis of data from the \hess, \magic, \astri \citep[e.g.][]{Vercellone2022}, and \veritas instruments.

While \gammapy was mainly developed for the science community around
IACT instruments, the internal data model and software design are general
enough to be applied to other \gammaray instruments as well.
The use of \gammapy for the analysis of data from the High Altitude
Water Cherenkov Observatory (HAWC) has been successfully
demonstrated by \cite{Olivera2022}. This makes \gammapy
a viable choice for the base library for the science tools
of the future Southern Widefield Gamma Ray Observatory
(SWGO) and use with data from the Large High Altitude Air Shower Observatory (LHAASO) as well. \gammapy
has the potential to further unify the community
of \gammaray astronomers, by sharing common tools, data formats, and
a common vision of open and reproducible science for the future.

\begin{acknowledgements}

        We would like to thank the \texttt{Numpy}, \texttt{Scipy}, \texttt{IPython} and
        \texttt{Matplotlib} communities for providing their packages which are
        invaluable to the development of \gammapy. We thank the \github team for
        providing us with an excellent free development platform. We also are grateful
        to Read the Docs (\ReadthedocsUrl), and Travis (\TravisUrl) for providing free
        documentation hosting and testing respectively. We would like to thank
        all the \gammapy users that have provided feedback and submitted bug reports.

        A.~Aguasca-Cabot acknowledges the financial support from the Spanish Ministry of Science and Innovation
        and the Spanish Research State Agency (AEI) under grant PID2019-104114RB-C33/AEI/10.13039/501100011034
        and the Institute of Cosmos Sciences University of Barcelona (ICCUB, Unidad de Excelencia “María de Maeztu”)
        through grant CEX2019-000918-M. J.L.~Contreras acknowledges the funding from the ESCAPE H2020 project, GA No 824064.
        L.~Giunti acknowledges financial support from the Agence Nationale de la Recherche (ANR-17-CE31-0014). M. Linhoff
        acknowledges support by the German BMBF (ErUM) and DFG (SFBs 876 and 1491).
        R.~López-Coto acknowledges the Ramon y Cajal program through grant RYC-2020-028639-I and the financial support
        from the grant CEX2021-001131-S funded by MCIN/AEI/ 10.13039/501100011033. C.~Nigro C.N. acknowledges support by
        the Spanish Ministerio de Ciencia e Innovación (MICINN), the European Union – NextGenerationEU and PRTR  through
        the programme Juan de la Cierva (grant FJC2020-046063-I), by the the MICINN (grant PID2019-107847RB-C41), and from
        the CERCA program of the Generalitat de Catalunya. Q.~Remy acknowledges support from the project "European Science
        Cluster of Astronomy \& Particle Physics ESFRI Research Infrastructures" (ESCAPE), that has received funding from
        the European Union’s Horizon 2020 research and innovation programme under Grant Agreement no. 824064. J.E.~Ruiz
        acknowledges financial support from the grant
        CEX2021-001131-S funded by MCIN/AEI/ 10.13039/501100011033. A.~Siemiginowska was supported by NASA contract
        NAS8-03060 (Chandra X-ray Center). A.~Sinha acknowledges support from The European Science Cluster of Astronomy \&
        Particle Physics ESFRI Research Infrastructures  funded by the European Union’s Horizon 2020 research and
        innovation program under Grant Agreement no. 824064 and from the Spanish Ministry of Universities through the
        Maria Zambrano Talent Attraction Programme, 2021-2023.

        A special acknowledgment has to be given to our first lead developer, Christoph Deil, who
        started the \gammapy project and set the foundation for its future success.

        For contributing to the writing of the manuscript, we thank A. Donath, R. Terrier, Q. Remy, A. Sinha,
        C. Nigro, F. Pintore, B. Khélifi, L. Olivera-Nieto, J. E. Ruiz, K. Brügge, M. Linhoff and J.L. Contreras.
        
        We also thank the current and former members of the Gammapy coordination committee A. Donath, B. Khélifi,
        C. Boisson, C.Deil, C. van Eldik, D. Berge, E. de Ona Wilhelmi, F. Acero, F. Pintore,
        M. Cardillo, J. Hinton, J.L. Contreras, M. Fuessling, R. Terrier, R. Zanin, R. López-Coto and S. Funk,
        who contributed to promotion, coordination and steering of the \gammapy project.

        Finally we would like to thank A. Aguasca-Cabot, P. Bhattacharjee, K. Brügge, J. Buchner, D. Carreto
        Fidalgo, A. Chen, M. de Bony de Lavergne, A. Donath, J. V. de Miranda Cardoso, C. Deil, L. Giunti,
        L. Jouvin, B. Khélifi, J. King, J. Lefaucheur, M. Lemoine-Goumard, J.P. Lenain, M. Linhoff,
        L. Mohrmann, D. Morcuende, C. Nigro, L. Olivera-Nieto, S. Panny, F. Pintore, M. Regeard,
        Q. Remy, J. E. Ruiz, L. Saha, H. Siejkowski, A. Siemiginowska, A. Sinha, B. M. Sipőcz, R. Terrier,
        T. Unbehaun, T. Vuillaume and unnamed contributors for contributing to the development of \gammapy.

\end{acknowledgements}

\bibliographystyle{aa}
\bibliography{bib}

\appendix
\section{Code examples' output}

\begin{figure}[!ht]
        \small
    	\import{code-examples/generated-output/}{gp_data}
        \caption{Output from the code example shown in Figure~\ref{fig*:minted:gp_data}.}
        \label{fig:code_example_gp_data}
\end{figure}

\begin{figure}[!ht]
        \small
        \import{code-examples/generated-output/}{gp_datasets}
        \caption{Output from the code example shown in Figure~\ref{fig*:minted:gp_datasets}.}
        \label{fig:code_example_gp_datasets}
\end{figure}

\begin{figure}[!ht]
        \small
        \import{code-examples/generated-output/}{gp_maps}
        \caption{Output from the code example shown in Figure~\ref{fig*:minted:gp_maps}.}
        \label{fig:code_example_gp_maps}
\end{figure}

\begin{figure}[!ht]
        \small
        \import{code-examples/generated-output/}{gp_stats}
        \caption{Output from the code example shown in Figure~\ref{fig*:minted:gp_stats}.}
        \label{fig:code_example_gp_stats}
\end{figure}

\begin{figure*}[!ht]
        \centering
        \includegraphics[width=1.\textwidth]{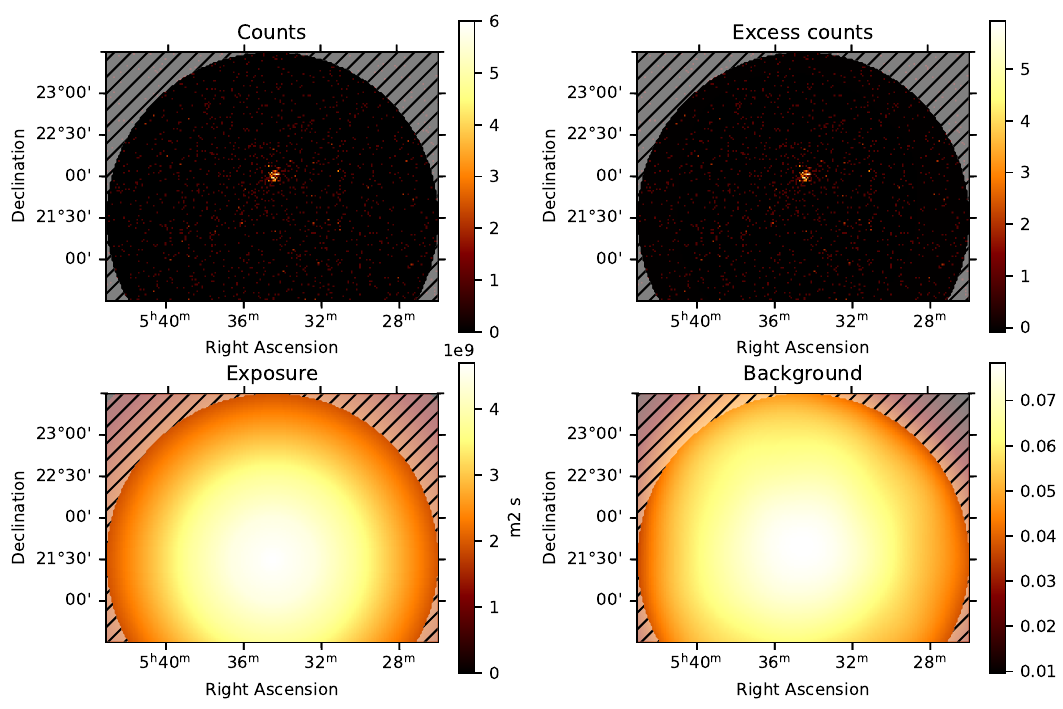}
        \caption{Output from the code example shown in Figure~\ref{fig*:minted:gp_makers}.}
        \label{fig:code_example_gp_makers}
\end{figure*}

\begin{figure*}[!ht]
        \centering
        \includegraphics[width=1.\textwidth]{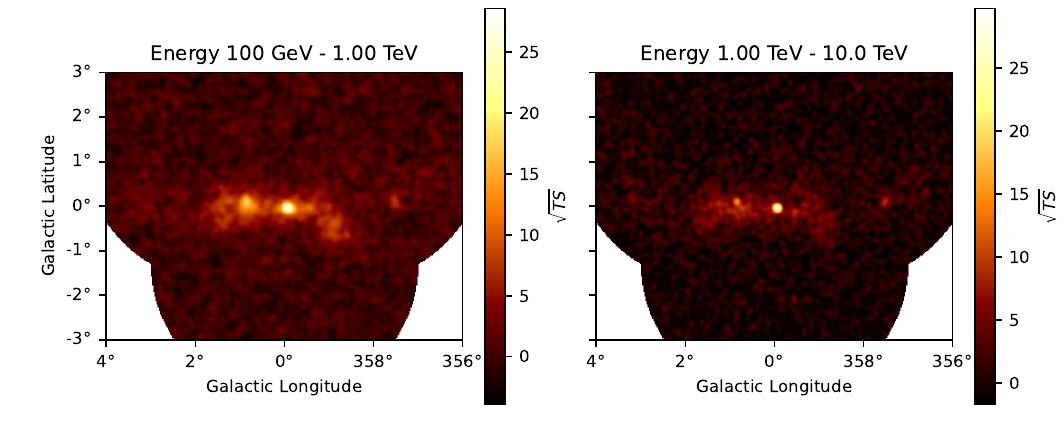}
        \caption{Output from the code example shown in Figure~\ref{fig*:minted:gp_estimators}.}
        \label{fig:code_example_gp_estimators}
\end{figure*}

\begin{figure*}[!ht]
        \centering
        \includegraphics[width=1.\textwidth]{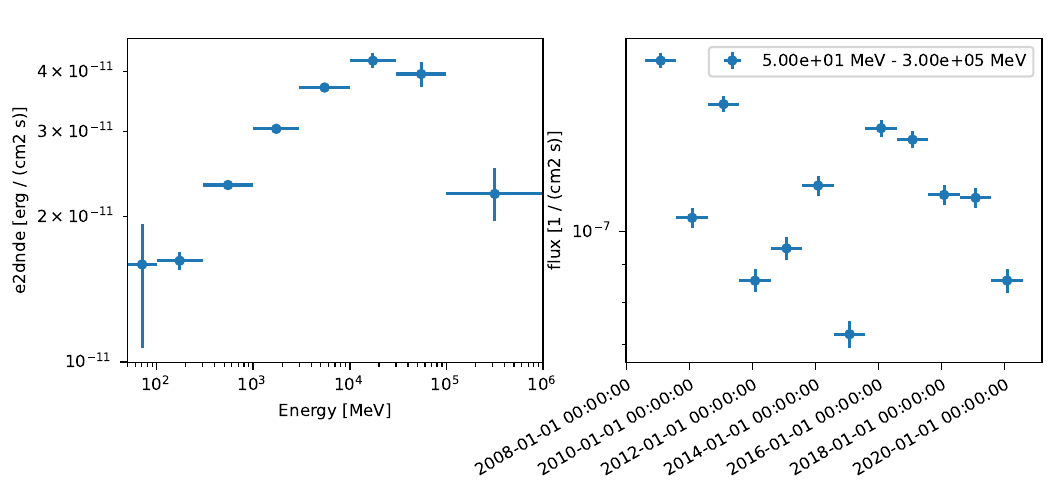}
        \caption{Output from the code example shown in Figure~\ref{fig*:minted:gp_catalogs}.}
        \label{fig:code_example_gp_catalogs}
\end{figure*}

\begin{figure}[!ht]
        \small
        \import{code-examples/generated-output/}{gp_models}
        \caption{Output from the code example shown in Figure~\ref{fig*:minted:gp_models}.}
        \label{fig:code_example_gp_models}
\end{figure}

\end{document}

%% file: code-examples/generated/gp_data.tex
\begin{Verbatim}[commandchars=\\\{\},frame=lines]
\PY{k+kn}{from} \PY{n+nn}{gammapy}\PY{n+nn}{.}\PY{n+nn}{data} \PY{k+kn}{import} \PY{n}{DataStore}

\PY{n}{data\PYZus{}store} \PY{o}{=} \PY{n}{DataStore}\PY{o}{.}\PY{n}{from\PYZus{}dir}\PY{p}{(}
    \PY{n}{base\PYZus{}dir}\PY{o}{=}\PY{l+s+s2}{\PYZdq{}}\PY{l+s+s2}{\PYZdl{}GAMMAPY\PYZus{}DATA/hess\PYZhy{}dl3\PYZhy{}dr1}\PY{l+s+s2}{\PYZdq{}}
\PY{p}{)}

\PY{n}{obs\PYZus{}ids} \PY{o}{=} \PY{p}{[}\PY{l+m+mi}{23523}\PY{p}{,} \PY{l+m+mi}{23526}\PY{p}{,} \PY{l+m+mi}{23559}\PY{p}{,} \PY{l+m+mi}{23592}\PY{p}{]}

\PY{n}{observations} \PY{o}{=} \PY{n}{data\PYZus{}store}\PY{o}{.}\PY{n}{get\PYZus{}observations}\PY{p}{(}
    \PY{n}{obs\PYZus{}id}\PY{o}{=}\PY{n}{obs\PYZus{}ids}\PY{p}{,} \PY{n}{skip\PYZus{}missing}\PY{o}{=}\PY{k+kc}{True}
\PY{p}{)}

\PY{k}{for} \PY{n}{obs} \PY{o+ow}{in} \PY{n}{observations}\PY{p}{:}
    \PY{n+nb}{print}\PY{p}{(}\PY{l+s+sa}{f}\PY{l+s+s2}{\PYZdq{}}\PY{l+s+s2}{Observation id: }\PY{l+s+si}{\PYZob{}}\PY{n}{obs}\PY{o}{.}\PY{n}{obs\PYZus{}id}\PY{l+s+si}{\PYZcb{}}\PY{l+s+s2}{\PYZdq{}}\PY{p}{)}
    \PY{n+nb}{print}\PY{p}{(}\PY{l+s+sa}{f}\PY{l+s+s2}{\PYZdq{}}\PY{l+s+s2}{N events: }\PY{l+s+si}{\PYZob{}}\PY{n+nb}{len}\PY{p}{(}\PY{n}{obs}\PY{o}{.}\PY{n}{events}\PY{o}{.}\PY{n}{table}\PY{p}{)}\PY{l+s+si}{\PYZcb{}}\PY{l+s+s2}{\PYZdq{}}\PY{p}{)}
    \PY{n+nb}{print}\PY{p}{(}\PY{l+s+sa}{f}\PY{l+s+s2}{\PYZdq{}}\PY{l+s+s2}{Max. area: }\PY{l+s+si}{\PYZob{}}\PY{n}{obs}\PY{o}{.}\PY{n}{aeff}\PY{o}{.}\PY{n}{quantity}\PY{o}{.}\PY{n}{max}\PY{p}{(}\PY{p}{)}\PY{l+s+si}{\PYZcb{}}\PY{l+s+s2}{\PYZdq{}}\PY{p}{)}
\end{Verbatim}

%% file: code-examples/generated/gp_maps.tex
\begin{Verbatim}[commandchars=\\\{\},frame=lines]
\PY{k+kn}{from} \PY{n+nn}{gammapy}\PY{n+nn}{.}\PY{n+nn}{maps} \PY{k+kn}{import} \PY{n}{Map}\PY{p}{,} \PY{n}{MapAxis}
\PY{k+kn}{from} \PY{n+nn}{astropy}\PY{n+nn}{.}\PY{n+nn}{coordinates} \PY{k+kn}{import} \PY{n}{SkyCoord}
\PY{k+kn}{from} \PY{n+nn}{astropy} \PY{k+kn}{import} \PY{n}{units} \PY{k}{as} \PY{n}{u}

\PY{n}{skydir} \PY{o}{=} \PY{n}{SkyCoord}\PY{p}{(}\PY{l+s+s2}{\PYZdq{}}\PY{l+s+s2}{0d}\PY{l+s+s2}{\PYZdq{}}\PY{p}{,} \PY{l+s+s2}{\PYZdq{}}\PY{l+s+s2}{5d}\PY{l+s+s2}{\PYZdq{}}\PY{p}{,} \PY{n}{frame}\PY{o}{=}\PY{l+s+s2}{\PYZdq{}}\PY{l+s+s2}{galactic}\PY{l+s+s2}{\PYZdq{}}\PY{p}{)}

\PY{n}{energy\PYZus{}axis} \PY{o}{=} \PY{n}{MapAxis}\PY{o}{.}\PY{n}{from\PYZus{}energy\PYZus{}bounds}\PY{p}{(}
    \PY{n}{energy\PYZus{}min}\PY{o}{=}\PY{l+s+s2}{\PYZdq{}}\PY{l+s+s2}{1 TeV}\PY{l+s+s2}{\PYZdq{}}\PY{p}{,} \PY{n}{energy\PYZus{}max}\PY{o}{=}\PY{l+s+s2}{\PYZdq{}}\PY{l+s+s2}{10 TeV}\PY{l+s+s2}{\PYZdq{}}\PY{p}{,} \PY{n}{nbin}\PY{o}{=}\PY{l+m+mi}{10}
\PY{p}{)}

\PY{c+c1}{\PYZsh{} Create a WCS Map}
\PY{n}{m\PYZus{}wcs} \PY{o}{=} \PY{n}{Map}\PY{o}{.}\PY{n}{create}\PY{p}{(}
    \PY{n}{binsz}\PY{o}{=}\PY{l+m+mf}{0.1}\PY{p}{,}
    \PY{n}{map\PYZus{}type}\PY{o}{=}\PY{l+s+s2}{\PYZdq{}}\PY{l+s+s2}{wcs}\PY{l+s+s2}{\PYZdq{}}\PY{p}{,}
    \PY{n}{skydir}\PY{o}{=}\PY{n}{skydir}\PY{p}{,}
    \PY{n}{width}\PY{o}{=}\PY{p}{[}\PY{l+m+mf}{10.0}\PY{p}{,} \PY{l+m+mf}{8.0}\PY{p}{]} \PY{o}{*} \PY{n}{u}\PY{o}{.}\PY{n}{deg}\PY{p}{,}
    \PY{n}{axes}\PY{o}{=}\PY{p}{[}\PY{n}{energy\PYZus{}axis}\PY{p}{]}\PY{p}{)}


\PY{c+c1}{\PYZsh{} Create a HEALPix Map}
\PY{n}{m\PYZus{}hpx} \PY{o}{=} \PY{n}{Map}\PY{o}{.}\PY{n}{create}\PY{p}{(}
    \PY{n}{binsz}\PY{o}{=}\PY{l+m+mf}{0.1}\PY{p}{,}
    \PY{n}{map\PYZus{}type}\PY{o}{=}\PY{l+s+s2}{\PYZdq{}}\PY{l+s+s2}{hpx}\PY{l+s+s2}{\PYZdq{}}\PY{p}{,}
    \PY{n}{skydir}\PY{o}{=}\PY{n}{skydir}\PY{p}{,}
    \PY{n}{axes}\PY{o}{=}\PY{p}{[}\PY{n}{energy\PYZus{}axis}\PY{p}{]}
\PY{p}{)}

\PY{c+c1}{\PYZsh{} Create a region map}
\PY{n}{region} \PY{o}{=} \PY{l+s+s2}{\PYZdq{}}\PY{l+s+s2}{galactic;circle(0, 5, 1)}\PY{l+s+s2}{\PYZdq{}}
\PY{n}{m\PYZus{}region} \PY{o}{=} \PY{n}{Map}\PY{o}{.}\PY{n}{create}\PY{p}{(}
    \PY{n}{region}\PY{o}{=}\PY{n}{region}\PY{p}{,}
    \PY{n}{map\PYZus{}type}\PY{o}{=}\PY{l+s+s2}{\PYZdq{}}\PY{l+s+s2}{region}\PY{l+s+s2}{\PYZdq{}}\PY{p}{,}
    \PY{n}{axes}\PY{o}{=}\PY{p}{[}\PY{n}{energy\PYZus{}axis}\PY{p}{]}
\PY{p}{)}

\PY{n+nb}{print}\PY{p}{(}\PY{n}{m\PYZus{}wcs}\PY{p}{,} \PY{n}{m\PYZus{}hpx}\PY{p}{,} \PY{n}{m\PYZus{}region}\PY{p}{)}
\end{Verbatim}

%% file: code-examples/generated/gp_datasets.tex
\begin{Verbatim}[commandchars=\\\{\},frame=lines]
\PY{k+kn}{from} \PY{n+nn}{pathlib} \PY{k+kn}{import} \PY{n}{Path}

\PY{k+kn}{from} \PY{n+nn}{gammapy}\PY{n+nn}{.}\PY{n+nn}{datasets} \PY{k+kn}{import} \PY{p}{(}
    \PY{n}{Datasets}\PY{p}{,}
    \PY{n}{FluxPointsDataset}\PY{p}{,}
    \PY{n}{MapDataset}\PY{p}{,}
    \PY{n}{SpectrumDatasetOnOff}\PY{p}{,}
\PY{p}{)}

\PY{n}{path} \PY{o}{=} \PY{n}{Path}\PY{p}{(}\PY{l+s+s2}{\PYZdq{}}\PY{l+s+s2}{\PYZdl{}GAMMAPY\PYZus{}DATA}\PY{l+s+s2}{\PYZdq{}}\PY{p}{)}

\PY{n}{map\PYZus{}dataset} \PY{o}{=} \PY{n}{MapDataset}\PY{o}{.}\PY{n}{read}\PY{p}{(}
    \PY{n}{path} \PY{o}{/} \PY{l+s+s2}{\PYZdq{}}\PY{l+s+s2}{cta\PYZhy{}1dc\PYZhy{}gc/cta\PYZhy{}1dc\PYZhy{}gc.fits.gz}\PY{l+s+s2}{\PYZdq{}}\PY{p}{,}
    \PY{n}{name}\PY{o}{=}\PY{l+s+s2}{\PYZdq{}}\PY{l+s+s2}{map\PYZhy{}dataset}\PY{l+s+s2}{\PYZdq{}}\PY{p}{,}
\PY{p}{)}

\PY{n}{spectrum\PYZus{}dataset} \PY{o}{=} \PY{n}{SpectrumDatasetOnOff}\PY{o}{.}\PY{n}{read}\PY{p}{(}
    \PY{n}{path} \PY{o}{/} \PY{l+s+s2}{\PYZdq{}}\PY{l+s+s2}{joint\PYZhy{}crab/spectra/hess/pha\PYZus{}obs23523.fits}\PY{l+s+s2}{\PYZdq{}}\PY{p}{,}
    \PY{n}{name}\PY{o}{=}\PY{l+s+s2}{\PYZdq{}}\PY{l+s+s2}{spectrum\PYZhy{}datasets}\PY{l+s+s2}{\PYZdq{}}\PY{p}{,}
\PY{p}{)}

\PY{n}{flux\PYZus{}points\PYZus{}dataset} \PY{o}{=} \PY{n}{FluxPointsDataset}\PY{o}{.}\PY{n}{read}\PY{p}{(}
    \PY{n}{path} \PY{o}{/} \PY{l+s+s2}{\PYZdq{}}\PY{l+s+s2}{hawc\PYZus{}crab/HAWC19\PYZus{}flux\PYZus{}points.fits}\PY{l+s+s2}{\PYZdq{}}\PY{p}{,}
    \PY{n}{name}\PY{o}{=}\PY{l+s+s2}{\PYZdq{}}\PY{l+s+s2}{flux\PYZhy{}points\PYZhy{}dataset}\PY{l+s+s2}{\PYZdq{}}\PY{p}{,}
\PY{p}{)}


\PY{n}{datasets} \PY{o}{=} \PY{n}{Datasets}\PY{p}{(}\PY{p}{[}
    \PY{n}{map\PYZus{}dataset}\PY{p}{,}
    \PY{n}{spectrum\PYZus{}dataset}\PY{p}{,}
    \PY{n}{flux\PYZus{}points\PYZus{}dataset}
\PY{p}{]}\PY{p}{)}

\PY{n+nb}{print}\PY{p}{(}\PY{n}{datasets}\PY{p}{[}\PY{l+s+s2}{\PYZdq{}}\PY{l+s+s2}{map\PYZhy{}dataset}\PY{l+s+s2}{\PYZdq{}}\PY{p}{]}\PY{p}{)}
\end{Verbatim}

%% file: code-examples/generated/gp_makers.tex
\begin{Verbatim}[commandchars=\\\{\},frame=lines]
\PY{k+kn}{import} \PY{n+nn}{astropy}\PY{n+nn}{.}\PY{n+nn}{units} \PY{k}{as} \PY{n+nn}{u}

\PY{k+kn}{from} \PY{n+nn}{gammapy}\PY{n+nn}{.}\PY{n+nn}{data} \PY{k+kn}{import} \PY{n}{DataStore}
\PY{k+kn}{from} \PY{n+nn}{gammapy}\PY{n+nn}{.}\PY{n+nn}{datasets} \PY{k+kn}{import} \PY{n}{MapDataset}
\PY{k+kn}{from} \PY{n+nn}{gammapy}\PY{n+nn}{.}\PY{n+nn}{makers} \PY{k+kn}{import} \PY{p}{(}
    \PY{n}{FoVBackgroundMaker}\PY{p}{,}
    \PY{n}{MapDatasetMaker}\PY{p}{,}
    \PY{n}{SafeMaskMaker}
\PY{p}{)}
\PY{k+kn}{from} \PY{n+nn}{gammapy}\PY{n+nn}{.}\PY{n+nn}{maps} \PY{k+kn}{import} \PY{n}{MapAxis}\PY{p}{,} \PY{n}{WcsGeom}

\PY{n}{data\PYZus{}store} \PY{o}{=} \PY{n}{DataStore}\PY{o}{.}\PY{n}{from\PYZus{}dir}\PY{p}{(}
    \PY{n}{base\PYZus{}dir}\PY{o}{=}\PY{l+s+s2}{\PYZdq{}}\PY{l+s+s2}{\PYZdl{}GAMMAPY\PYZus{}DATA/hess\PYZhy{}dl3\PYZhy{}dr1}\PY{l+s+s2}{\PYZdq{}}
\PY{p}{)}

\PY{n}{obs} \PY{o}{=} \PY{n}{data\PYZus{}store}\PY{o}{.}\PY{n}{obs}\PY{p}{(}\PY{l+m+mi}{23523}\PY{p}{)}

\PY{n}{energy\PYZus{}axis} \PY{o}{=} \PY{n}{MapAxis}\PY{o}{.}\PY{n}{from\PYZus{}energy\PYZus{}bounds}\PY{p}{(}
    \PY{n}{energy\PYZus{}min}\PY{o}{=}\PY{l+s+s2}{\PYZdq{}}\PY{l+s+s2}{1 TeV}\PY{l+s+s2}{\PYZdq{}}\PY{p}{,}
    \PY{n}{energy\PYZus{}max}\PY{o}{=}\PY{l+s+s2}{\PYZdq{}}\PY{l+s+s2}{10 TeV}\PY{l+s+s2}{\PYZdq{}}\PY{p}{,}
    \PY{n}{nbin}\PY{o}{=}\PY{l+m+mi}{6}\PY{p}{,}
\PY{p}{)}

\PY{n}{geom} \PY{o}{=} \PY{n}{WcsGeom}\PY{o}{.}\PY{n}{create}\PY{p}{(}
    \PY{n}{skydir}\PY{o}{=}\PY{p}{(}\PY{l+m+mf}{83.633}\PY{p}{,} \PY{l+m+mf}{22.014}\PY{p}{)}\PY{p}{,}
    \PY{n}{width}\PY{o}{=}\PY{p}{(}\PY{l+m+mi}{4}\PY{p}{,} \PY{l+m+mi}{3}\PY{p}{)} \PY{o}{*} \PY{n}{u}\PY{o}{.}\PY{n}{deg}\PY{p}{,}
    \PY{n}{axes}\PY{o}{=}\PY{p}{[}\PY{n}{energy\PYZus{}axis}\PY{p}{]}\PY{p}{,}
    \PY{n}{binsz}\PY{o}{=}\PY{l+m+mf}{0.02} \PY{o}{*} \PY{n}{u}\PY{o}{.}\PY{n}{deg}\PY{p}{,}
\PY{p}{)}

\PY{n}{empty} \PY{o}{=} \PY{n}{MapDataset}\PY{o}{.}\PY{n}{create}\PY{p}{(}\PY{n}{geom}\PY{o}{=}\PY{n}{geom}\PY{p}{)}

\PY{n}{maker} \PY{o}{=} \PY{n}{MapDatasetMaker}\PY{p}{(}\PY{p}{)}

\PY{n}{mask\PYZus{}maker} \PY{o}{=} \PY{n}{SafeMaskMaker}\PY{p}{(}
    \PY{n}{methods}\PY{o}{=}\PY{p}{[}\PY{l+s+s2}{\PYZdq{}}\PY{l+s+s2}{offset\PYZhy{}max}\PY{l+s+s2}{\PYZdq{}}\PY{p}{,} \PY{l+s+s2}{\PYZdq{}}\PY{l+s+s2}{aeff\PYZhy{}default}\PY{l+s+s2}{\PYZdq{}}\PY{p}{]}\PY{p}{,}
    \PY{n}{offset\PYZus{}max}\PY{o}{=}\PY{l+s+s2}{\PYZdq{}}\PY{l+s+s2}{2.0 deg}\PY{l+s+s2}{\PYZdq{}}\PY{p}{,}
\PY{p}{)}

\PY{n}{bkg\PYZus{}maker} \PY{o}{=} \PY{n}{FoVBackgroundMaker}\PY{p}{(}
    \PY{n}{method}\PY{o}{=}\PY{l+s+s2}{\PYZdq{}}\PY{l+s+s2}{scale}\PY{l+s+s2}{\PYZdq{}}\PY{p}{,}
\PY{p}{)}

\PY{n}{dataset} \PY{o}{=} \PY{n}{maker}\PY{o}{.}\PY{n}{run}\PY{p}{(}\PY{n}{empty}\PY{p}{,} \PY{n}{observation}\PY{o}{=}\PY{n}{obs}\PY{p}{)}
\PY{n}{dataset} \PY{o}{=} \PY{n}{bkg\PYZus{}maker}\PY{o}{.}\PY{n}{run}\PY{p}{(}\PY{n}{dataset}\PY{p}{,} \PY{n}{observation}\PY{o}{=}\PY{n}{obs}\PY{p}{)}
\PY{n}{dataset} \PY{o}{=} \PY{n}{mask\PYZus{}maker}\PY{o}{.}\PY{n}{run}\PY{p}{(}\PY{n}{dataset}\PY{p}{,} \PY{n}{observation}\PY{o}{=}\PY{n}{obs}\PY{p}{)}
\PY{n}{dataset}\PY{o}{.}\PY{n}{peek}\PY{p}{(}\PY{p}{)}
\end{Verbatim}

%% file: code-examples/generated/gp_stats.tex
\begin{Verbatim}[commandchars=\\\{\},frame=lines]
\PY{k+kn}{from} \PY{n+nn}{gammapy}\PY{n+nn}{.}\PY{n+nn}{stats} \PY{k+kn}{import} \PY{n}{WStatCountsStatistic}

\PY{n}{n\PYZus{}on} \PY{o}{=} \PY{p}{[}\PY{l+m+mi}{13}\PY{p}{,} \PY{l+m+mi}{5}\PY{p}{,} \PY{l+m+mi}{3}\PY{p}{]}
\PY{n}{n\PYZus{}off} \PY{o}{=} \PY{p}{[}\PY{l+m+mi}{11}\PY{p}{,} \PY{l+m+mi}{9}\PY{p}{,} \PY{l+m+mi}{20}\PY{p}{]}
\PY{n}{alpha} \PY{o}{=} \PY{p}{[}\PY{l+m+mf}{0.8}\PY{p}{,} \PY{l+m+mf}{0.5}\PY{p}{,} \PY{l+m+mf}{0.1}\PY{p}{]}
\PY{n}{stat} \PY{o}{=} \PY{n}{WStatCountsStatistic}\PY{p}{(}\PY{n}{n\PYZus{}on}\PY{p}{,} \PY{n}{n\PYZus{}off}\PY{p}{,} \PY{n}{alpha}\PY{p}{)}

\PY{c+c1}{\PYZsh{} Excess}
\PY{n+nb}{print}\PY{p}{(}\PY{l+s+sa}{f}\PY{l+s+s2}{\PYZdq{}}\PY{l+s+s2}{Excess: }\PY{l+s+si}{\PYZob{}}\PY{n}{stat}\PY{o}{.}\PY{n}{n\PYZus{}sig}\PY{l+s+si}{\PYZcb{}}\PY{l+s+s2}{\PYZdq{}}\PY{p}{)}

\PY{c+c1}{\PYZsh{} Significance}
\PY{n+nb}{print}\PY{p}{(}\PY{l+s+sa}{f}\PY{l+s+s2}{\PYZdq{}}\PY{l+s+s2}{Significance: }\PY{l+s+si}{\PYZob{}}\PY{n}{stat}\PY{o}{.}\PY{n}{sqrt\PYZus{}ts}\PY{l+s+si}{\PYZcb{}}\PY{l+s+s2}{\PYZdq{}}\PY{p}{)}

\PY{c+c1}{\PYZsh{} Asymmetrical errors}
\PY{n+nb}{print}\PY{p}{(}\PY{l+s+sa}{f}\PY{l+s+s2}{\PYZdq{}}\PY{l+s+s2}{Error Neg.: }\PY{l+s+si}{\PYZob{}}\PY{n}{stat}\PY{o}{.}\PY{n}{compute\PYZus{}errn}\PY{p}{(}\PY{n}{n\PYZus{}sigma}\PY{o}{=}\PY{l+m+mf}{1.0}\PY{p}{)}\PY{l+s+si}{\PYZcb{}}\PY{l+s+s2}{\PYZdq{}}\PY{p}{)}
\PY{n+nb}{print}\PY{p}{(}\PY{l+s+sa}{f}\PY{l+s+s2}{\PYZdq{}}\PY{l+s+s2}{Error Pos.: }\PY{l+s+si}{\PYZob{}}\PY{n}{stat}\PY{o}{.}\PY{n}{compute\PYZus{}errp}\PY{p}{(}\PY{n}{n\PYZus{}sigma}\PY{o}{=}\PY{l+m+mf}{1.0}\PY{p}{)}\PY{l+s+si}{\PYZcb{}}\PY{l+s+s2}{\PYZdq{}}\PY{p}{)}
\end{Verbatim}

%% file: code-examples/generated/gp_models.tex
\begin{Verbatim}[commandchars=\\\{\},frame=lines]
\PY{k+kn}{from} \PY{n+nn}{astropy} \PY{k+kn}{import} \PY{n}{units} \PY{k}{as} \PY{n}{u}
\PY{k+kn}{from} \PY{n+nn}{gammapy}\PY{n+nn}{.}\PY{n+nn}{modeling}\PY{n+nn}{.}\PY{n+nn}{models} \PY{k+kn}{import} \PY{p}{(}
    \PY{n}{ConstantTemporalModel}\PY{p}{,}
    \PY{n}{EBLAbsorptionNormSpectralModel}\PY{p}{,}
    \PY{n}{PointSpatialModel}\PY{p}{,}
    \PY{n}{PowerLawSpectralModel}\PY{p}{,}
    \PY{n}{SkyModel}\PY{p}{,}
\PY{p}{)}

\PY{c+c1}{\PYZsh{} define a spectral model}
\PY{n}{pwl} \PY{o}{=} \PY{n}{PowerLawSpectralModel}\PY{p}{(}
    \PY{n}{amplitude}\PY{o}{=}\PY{l+s+s2}{\PYZdq{}}\PY{l+s+s2}{1e\PYZhy{}12 TeV\PYZhy{}1 cm\PYZhy{}2 s\PYZhy{}1}\PY{l+s+s2}{\PYZdq{}}\PY{p}{,} \PY{n}{index}\PY{o}{=}\PY{l+m+mf}{2.3}
\PY{p}{)}

\PY{c+c1}{\PYZsh{} define a spatial model}
\PY{n}{point} \PY{o}{=} \PY{n}{PointSpatialModel}\PY{p}{(}
    \PY{n}{lon\PYZus{}0}\PY{o}{=}\PY{l+s+s2}{\PYZdq{}}\PY{l+s+s2}{45.6 deg}\PY{l+s+s2}{\PYZdq{}}\PY{p}{,}
    \PY{n}{lat\PYZus{}0}\PY{o}{=}\PY{l+s+s2}{\PYZdq{}}\PY{l+s+s2}{3.2 deg}\PY{l+s+s2}{\PYZdq{}}\PY{p}{,}
    \PY{n}{frame}\PY{o}{=}\PY{l+s+s2}{\PYZdq{}}\PY{l+s+s2}{galactic}\PY{l+s+s2}{\PYZdq{}}
\PY{p}{)}


\PY{c+c1}{\PYZsh{} define a temporal model}
\PY{n}{constant} \PY{o}{=} \PY{n}{ConstantTemporalModel}\PY{p}{(}\PY{p}{)}

\PY{c+c1}{\PYZsh{} combine all components}
\PY{n}{model} \PY{o}{=} \PY{n}{SkyModel}\PY{p}{(}
    \PY{n}{spectral\PYZus{}model}\PY{o}{=}\PY{n}{pwl}\PY{p}{,}
    \PY{n}{spatial\PYZus{}model}\PY{o}{=}\PY{n}{point}\PY{p}{,}
    \PY{n}{temporal\PYZus{}model}\PY{o}{=}\PY{n}{constant}\PY{p}{,}
    \PY{n}{name}\PY{o}{=}\PY{l+s+s2}{\PYZdq{}}\PY{l+s+s2}{my\PYZhy{}model}\PY{l+s+s2}{\PYZdq{}}\PY{p}{,}
\PY{p}{)}
\PY{n+nb}{print}\PY{p}{(}\PY{n}{model}\PY{p}{)}

\PY{n}{ebl} \PY{o}{=} \PY{n}{EBLAbsorptionNormSpectralModel}\PY{o}{.}\PY{n}{read\PYZus{}builtin}\PY{p}{(}
    \PY{n}{reference}\PY{o}{=}\PY{l+s+s2}{\PYZdq{}}\PY{l+s+s2}{dominguez}\PY{l+s+s2}{\PYZdq{}}\PY{p}{,} \PY{n}{redshift}\PY{o}{=}\PY{l+m+mf}{0.5}
\PY{p}{)}

\PY{n}{absorbed} \PY{o}{=} \PY{n}{pwl} \PY{o}{*} \PY{n}{ebl}
\PY{n}{absorbed}\PY{o}{.}\PY{n}{plot}\PY{p}{(}\PY{n}{energy\PYZus{}bounds}\PY{o}{=}\PY{p}{(}\PY{l+m+mf}{0.1}\PY{p}{,} \PY{l+m+mi}{100}\PY{p}{)} \PY{o}{*} \PY{n}{u}\PY{o}{.}\PY{n}{TeV}\PY{p}{)}
\end{Verbatim}

%% file: code-examples/generated/gp_estimators.tex
\begin{Verbatim}[commandchars=\\\{\},frame=lines]
\PY{k+kn}{from} \PY{n+nn}{astropy} \PY{k+kn}{import} \PY{n}{units} \PY{k}{as} \PY{n}{u}

\PY{k+kn}{from} \PY{n+nn}{gammapy}\PY{n+nn}{.}\PY{n+nn}{datasets} \PY{k+kn}{import} \PY{n}{MapDataset}
\PY{k+kn}{from} \PY{n+nn}{gammapy}\PY{n+nn}{.}\PY{n+nn}{estimators} \PY{k+kn}{import} \PY{n}{TSMapEstimator}

\PY{n}{filename} \PY{o}{=} \PY{l+s+s2}{\PYZdq{}}\PY{l+s+s2}{\PYZdl{}GAMMAPY\PYZus{}DATA/cta\PYZhy{}1dc\PYZhy{}gc/cta\PYZhy{}1dc\PYZhy{}gc.fits.gz}\PY{l+s+s2}{\PYZdq{}}
\PY{n}{dataset} \PY{o}{=} \PY{n}{MapDataset}\PY{o}{.}\PY{n}{read}\PY{p}{(}\PY{n}{filename}\PY{p}{)}

\PY{n}{estimator} \PY{o}{=} \PY{n}{TSMapEstimator}\PY{p}{(}
    \PY{n}{energy\PYZus{}edges}\PY{o}{=}\PY{p}{[}\PY{l+m+mf}{0.1}\PY{p}{,} \PY{l+m+mi}{1}\PY{p}{,} \PY{l+m+mi}{10}\PY{p}{]} \PY{o}{*} \PY{n}{u}\PY{o}{.}\PY{n}{TeV}\PY{p}{,}
    \PY{n}{n\PYZus{}sigma}\PY{o}{=}\PY{l+m+mi}{1}\PY{p}{,}
    \PY{n}{n\PYZus{}sigma\PYZus{}ul}\PY{o}{=}\PY{l+m+mi}{2}\PY{p}{,}
\PY{p}{)}

\PY{n}{maps} \PY{o}{=} \PY{n}{estimator}\PY{o}{.}\PY{n}{run}\PY{p}{(}\PY{n}{dataset}\PY{p}{)}
\PY{n}{maps}\PY{p}{[}\PY{l+s+s2}{\PYZdq{}}\PY{l+s+s2}{sqrt\PYZus{}ts}\PY{l+s+s2}{\PYZdq{}}\PY{p}{]}\PY{o}{.}\PY{n}{plot\PYZus{}grid}\PY{p}{(}\PY{n}{add\PYZus{}cbar}\PY{o}{=}\PY{k+kc}{True}\PY{p}{)}
\end{Verbatim}

%% file: code-examples/generated/gp_catalogs.tex
\begin{Verbatim}[commandchars=\\\{\},frame=lines]
\PY{k+kn}{import} \PY{n+nn}{matplotlib}\PY{n+nn}{.}\PY{n+nn}{pyplot} \PY{k}{as} \PY{n+nn}{plt}

\PY{k+kn}{from} \PY{n+nn}{gammapy}\PY{n+nn}{.}\PY{n+nn}{catalog} \PY{k+kn}{import} \PY{n}{CATALOG\PYZus{}REGISTRY}

\PY{n}{catalog} \PY{o}{=} \PY{n}{CATALOG\PYZus{}REGISTRY}\PY{o}{.}\PY{n}{get\PYZus{}cls}\PY{p}{(}\PY{l+s+s2}{\PYZdq{}}\PY{l+s+s2}{4fgl}\PY{l+s+s2}{\PYZdq{}}\PY{p}{)}\PY{p}{(}\PY{p}{)}
\PY{n+nb}{print}\PY{p}{(}\PY{l+s+s2}{\PYZdq{}}\PY{l+s+s2}{Number of sources :}\PY{l+s+s2}{\PYZdq{}}\PY{p}{,} \PY{n+nb}{len}\PY{p}{(}\PY{n}{catalog}\PY{o}{.}\PY{n}{table}\PY{p}{)}\PY{p}{)}

\PY{n}{source} \PY{o}{=} \PY{n}{catalog}\PY{p}{[}\PY{l+s+s2}{\PYZdq{}}\PY{l+s+s2}{PKS 2155\PYZhy{}304}\PY{l+s+s2}{\PYZdq{}}\PY{p}{]}

\PY{n}{\PYZus{}}\PY{p}{,} \PY{n}{axes} \PY{o}{=} \PY{n}{plt}\PY{o}{.}\PY{n}{subplots}\PY{p}{(}\PY{n}{ncols}\PY{o}{=}\PY{l+m+mi}{2}\PY{p}{)}
\PY{n}{source}\PY{o}{.}\PY{n}{flux\PYZus{}points}\PY{o}{.}\PY{n}{plot}\PY{p}{(}\PY{n}{ax}\PY{o}{=}\PY{n}{axes}\PY{p}{[}\PY{l+m+mi}{0}\PY{p}{]}\PY{p}{,} \PY{n}{sed\PYZus{}type}\PY{o}{=}\PY{l+s+s2}{\PYZdq{}}\PY{l+s+s2}{e2dnde}\PY{l+s+s2}{\PYZdq{}}\PY{p}{)}

\PY{n}{source}\PY{o}{.}\PY{n}{lightcurve}\PY{p}{(}\PY{p}{)}\PY{o}{.}\PY{n}{plot}\PY{p}{(}\PY{n}{ax}\PY{o}{=}\PY{n}{axes}\PY{p}{[}\PY{l+m+mi}{1}\PY{p}{]}\PY{p}{)}
\end{Verbatim}

%% file: code-examples/generated-output/gp_data.tex
\begin{Verbatim}[commandchars=\\\{\},frame=lines]
Observation id: 23523
N events: 7613
Max. area: 699771.0625 m2
Observation id: 23526
N events: 7581
Max. area: 623679.5 m2
Observation id: 23559
N events: 7601
Max. area: 613097.6875 m2
Observation id: 23592
N events: 7334
Max. area: 693575.75 m2
\end{Verbatim}

%% file: code-examples/generated-output/gp_datasets.tex
\begin{Verbatim}[commandchars=\\\{\},frame=lines]
\PY{g+gh}{MapDataset}
\PY{g+gh}{\PYZhy{}\PYZhy{}\PYZhy{}\PYZhy{}\PYZhy{}\PYZhy{}\PYZhy{}\PYZhy{}\PYZhy{}\PYZhy{}}

  Name                            : map\PYZhy{}dataset 

  Total counts                    : 104317 
  Total background counts         : 91507.70
  Total excess counts             : 12809.30

  Predicted counts                : 91507.69
  Predicted background counts     : 91507.70
  Predicted excess counts         : nan

  Exposure min                    : 6.28e+07 m2 s
  Exposure max                    : 1.90e+10 m2 s

  Number of total bins            : 768000 
  Number of fit bins              : 691680 

  Fit statistic type              : cash
  Fit statistic value (\PYZhy{}2 log(L)) : nan

  Number of models                : 0 
  Number of parameters            : 0
  Number of free parameters       : 0
\end{Verbatim}

%% file: code-examples/generated-output/gp_maps.tex
\begin{Verbatim}[commandchars=\\\{\},frame=lines]
\PY{n+nx}{WcsNDMap}

\PY{+w}{	}\PY{n+nx}{geom}\PY{+w}{  }\PY{p}{:}\PY{+w}{ }\PY{n+nx}{WcsGeom}\PY{+w}{ }
\PY{+w}{ 	}\PY{n+nx}{axes}\PY{+w}{  }\PY{p}{:}\PY{+w}{ }\PY{p}{[}\PY{err}{\PYZsq{}}\PY{n+nx}{lon}\PY{err}{\PYZsq{}}\PY{p}{,}\PY{+w}{ }\PY{err}{\PYZsq{}}\PY{n+nx}{lat}\PY{err}{\PYZsq{}}\PY{p}{,}\PY{+w}{ }\PY{err}{\PYZsq{}}\PY{n+nx}{energy}\PY{err}{\PYZsq{}}\PY{p}{]}
\PY{+w}{	}\PY{n+nx}{shape}\PY{+w}{ }\PY{p}{:}\PY{+w}{ }\PY{p}{(}\PY{l+m+mi}{1}\PY{l+m+mi}{0}\PY{l+m+mi}{0}\PY{p}{,}\PY{+w}{ }\PY{l+m+mi}{8}\PY{l+m+mi}{0}\PY{p}{,}\PY{+w}{ }\PY{l+m+mi}{1}\PY{l+m+mi}{0}\PY{p}{)}
\PY{+w}{	}\PY{n+nx}{ndim}\PY{+w}{  }\PY{p}{:}\PY{+w}{ }\PY{l+m+mi}{3}
\PY{+w}{	}\PY{n+nx}{unit}\PY{+w}{  }\PY{p}{:}\PY{+w}{ }
\PY{+w}{	}\PY{n+nx}{dtype}\PY{+w}{ }\PY{p}{:}\PY{+w}{ }\PY{n+nx}{float32}
\PY{+w}{ }\PY{n+nx}{HpxNDMap}

\PY{+w}{	}\PY{n+nx}{geom}\PY{+w}{  }\PY{p}{:}\PY{+w}{ }\PY{n+nx}{HpxGeom}\PY{+w}{ }
\PY{+w}{ 	}\PY{n+nx}{axes}\PY{+w}{  }\PY{p}{:}\PY{+w}{ }\PY{p}{[}\PY{err}{\PYZsq{}}\PY{n+nx}{skycoord}\PY{err}{\PYZsq{}}\PY{p}{,}\PY{+w}{ }\PY{err}{\PYZsq{}}\PY{n+nx}{energy}\PY{err}{\PYZsq{}}\PY{p}{]}
\PY{+w}{	}\PY{n+nx}{shape}\PY{+w}{ }\PY{p}{:}\PY{+w}{ }\PY{p}{(}\PY{l+m+mi}{3}\PY{l+m+mi}{1}\PY{l+m+mi}{4}\PY{l+m+mi}{5}\PY{l+m+mi}{7}\PY{l+m+mi}{2}\PY{l+m+mi}{8}\PY{p}{,}\PY{+w}{ }\PY{l+m+mi}{1}\PY{l+m+mi}{0}\PY{p}{)}
\PY{+w}{	}\PY{n+nx}{ndim}\PY{+w}{  }\PY{p}{:}\PY{+w}{ }\PY{l+m+mi}{3}
\PY{+w}{	}\PY{n+nx}{unit}\PY{+w}{  }\PY{p}{:}\PY{+w}{ }
\PY{+w}{	}\PY{n+nx}{dtype}\PY{+w}{ }\PY{p}{:}\PY{+w}{ }\PY{n+nx}{float32}
\PY{+w}{ }\PY{n+nx}{RegionNDMap}

\PY{+w}{	}\PY{n+nx}{geom}\PY{+w}{  }\PY{p}{:}\PY{+w}{ }\PY{n+nx}{RegionGeom}\PY{+w}{ }
\PY{+w}{ 	}\PY{n+nx}{axes}\PY{+w}{  }\PY{p}{:}\PY{+w}{ }\PY{p}{[}\PY{err}{\PYZsq{}}\PY{n+nx}{lon}\PY{err}{\PYZsq{}}\PY{p}{,}\PY{+w}{ }\PY{err}{\PYZsq{}}\PY{n+nx}{lat}\PY{err}{\PYZsq{}}\PY{p}{,}\PY{+w}{ }\PY{err}{\PYZsq{}}\PY{n+nx}{energy}\PY{err}{\PYZsq{}}\PY{p}{]}
\PY{+w}{	}\PY{n+nx}{shape}\PY{+w}{ }\PY{p}{:}\PY{+w}{ }\PY{p}{(}\PY{l+m+mi}{1}\PY{p}{,}\PY{+w}{ }\PY{l+m+mi}{1}\PY{p}{,}\PY{+w}{ }\PY{l+m+mi}{1}\PY{l+m+mi}{0}\PY{p}{)}
\PY{+w}{	}\PY{n+nx}{ndim}\PY{+w}{  }\PY{p}{:}\PY{+w}{ }\PY{l+m+mi}{3}
\PY{+w}{	}\PY{n+nx}{unit}\PY{+w}{  }\PY{p}{:}\PY{+w}{ }
\PY{+w}{	}\PY{n+nx}{dtype}\PY{+w}{ }\PY{p}{:}\PY{+w}{ }\PY{n+nx}{float32}
\end{Verbatim}

%% file: code-examples/generated-output/gp_stats.tex
\begin{Verbatim}[commandchars=\\\{\},frame=lines]
Excess: [4.2 0.5 1. ]
Significance: [0.95461389 0.18791253 0.62290414]
Error Neg.: [4.3980796  2.56480097 1.50533827]
Error Pos.: [4.63826301 2.91371256 2.11988712]
\end{Verbatim}

%% file: code-examples/generated-output/gp_models.tex
\begin{Verbatim}[commandchars=\\\{\},frame=lines]
\PY{n+nx}{SkyModel}

\PY{+w}{  }\PY{n+nx}{Name}\PY{+w}{                      }\PY{p}{:}\PY{+w}{ }\PY{n+nx}{my}\PY{o}{\PYZhy{}}\PY{n+nx}{model}
\PY{+w}{  }\PY{n+nx}{Datasets}\PY{+w}{ }\PY{n+nx}{names}\PY{+w}{            }\PY{p}{:}\PY{+w}{ }\PY{n+nx}{None}
\PY{+w}{  }\PY{n+nx}{Spectral}\PY{+w}{ }\PY{n+nx}{model}\PY{+w}{ }\PY{k}{type}\PY{+w}{       }\PY{p}{:}\PY{+w}{ }\PY{n+nx}{PowerLawSpectralModel}
\PY{+w}{  }\PY{n+nx}{Spatial}\PY{+w}{  }\PY{n+nx}{model}\PY{+w}{ }\PY{k}{type}\PY{+w}{       }\PY{p}{:}\PY{+w}{ }\PY{n+nx}{PointSpatialModel}
\PY{+w}{  }\PY{n+nx}{Temporal}\PY{+w}{ }\PY{n+nx}{model}\PY{+w}{ }\PY{k}{type}\PY{+w}{       }\PY{p}{:}\PY{+w}{ }\PY{n+nx}{ConstantTemporalModel}
\PY{+w}{  }\PY{n+nx}{Parameters}\PY{p}{:}
\PY{+w}{    }\PY{n+nx}{index}\PY{+w}{                         }\PY{p}{:}\PY{+w}{      }\PY{l+m+mDouble}{2.3}\PY{l+m+mi}{0}\PY{l+m+mi}{0}\PY{+w}{   }\PY{o}{+}\PY{o}{/}\PY{o}{\PYZhy{}}\PY{+w}{    }\PY{l+m+mDouble}{0.0}\PY{l+m+mi}{0}\PY{+w}{             }
\PY{+w}{    }\PY{n+nx}{amplitude}\PY{+w}{                     }\PY{p}{:}\PY{+w}{   }\PY{l+m+mDouble}{1.0}\PY{l+m+mi}{0}\PY{n+nx}{e}\PY{o}{\PYZhy{}}\PY{l+m+mi}{1}\PY{l+m+mi}{2}\PY{+w}{   }\PY{o}{+}\PY{o}{/}\PY{o}{\PYZhy{}}\PY{+w}{ }\PY{l+m+mDouble}{0.0}\PY{n+nx}{e}\PY{o}{+}\PY{l+m+mi}{0}\PY{l+m+mi}{0}\PY{+w}{ }\PY{l+m+mi}{1}\PY{+w}{ }\PY{o}{/}\PY{+w}{ }\PY{p}{(}\PY{n+nx}{cm2}\PY{+w}{ }\PY{n+nx}{s}\PY{+w}{ }\PY{n+nx}{TeV}\PY{p}{)}
\PY{+w}{    }\PY{n+nx}{reference}\PY{+w}{             }\PY{p}{(}\PY{n+nx}{frozen}\PY{p}{)}\PY{p}{:}\PY{+w}{      }\PY{l+m+mDouble}{1.0}\PY{l+m+mi}{0}\PY{l+m+mi}{0}\PY{+w}{       }\PY{n+nx}{TeV}\PY{+w}{         }
\PY{+w}{    }\PY{n+nx}{lon\PYZus{}0}\PY{+w}{                         }\PY{p}{:}\PY{+w}{     }\PY{l+m+mDouble}{45.6}\PY{l+m+mi}{0}\PY{l+m+mi}{0}\PY{+w}{   }\PY{o}{+}\PY{o}{/}\PY{o}{\PYZhy{}}\PY{+w}{    }\PY{l+m+mDouble}{0.0}\PY{l+m+mi}{0}\PY{+w}{ }\PY{n+nx}{deg}\PY{+w}{         }
\PY{+w}{    }\PY{n+nx}{lat\PYZus{}0}\PY{+w}{                         }\PY{p}{:}\PY{+w}{      }\PY{l+m+mDouble}{3.2}\PY{l+m+mi}{0}\PY{l+m+mi}{0}\PY{+w}{   }\PY{o}{+}\PY{o}{/}\PY{o}{\PYZhy{}}\PY{+w}{    }\PY{l+m+mDouble}{0.0}\PY{l+m+mi}{0}\PY{+w}{ }\PY{n+nx}{deg}\PY{+w}{         }
\end{Verbatim}